\documentclass[onecolumn,preprintnumbers,amsmath,prc,amssymb]{revtex4}
\usepackage{graphicx}
\usepackage{bm}
 
\begin{document}

\title{The statistical multifragmentation model for liquid-gas phase transition  with a compressible  nuclear liquid}

\author{V. V. Sagun$^1$, A. I. Ivanytskyi$^1$, K. A. Bugaev$^{1,2}$ and I. N. Mishustin$^{2,3}$}

\vspace*{11mm}

\affiliation{$^1$Bogolyubov Institute for Theoretical Physics of the National Academy of Sciences of Ukraine,\\ Metrologichna str. 14$^b$, Kiev-03680, Ukraine}

\affiliation{$^2$Frankfurt Institute for Advanced Studies (FIAS),
Goethe-University,  Ruth-Moufang Str. 1, 60438 Frankfurt upon Main, Germany}

\affiliation{$^3$ Kurchatov Institute, Russian Research Center, Akademika Kurchatova Sqr., Moscow, 123182, Russia}



\large 

\begin{abstract}
\large
We  propose a new formulation of the statistical multifragmentation model  based on the analysis 
of the virial expansion  for a system  of  the nuclear fragments of all sizes.
The developed model not only enables  us to   account for short-range repulsion, but also  to 
calculate  the surface  free energy which is  induced 
by the interaction between the fragments. 
  We propose  a new  parameterization for the liquid  phase pressure which  allows us to introduce   a compressible nuclear liquid  into  the statistical multifragmentation model. 
The resulting  model is exactly solvable and has no irregular behavior of the isotherms in the mixed phase region that  is typical 
for mean-field  models.  The general conditions for the 1-st  and   2-nd (or higher) order phase transitions are formulated.  It is shown that all  endpoints of the present model phase diagram are 
  the tricritical points, if    the Fisher exponent $\tau$ is  in the range   $\frac{3}{2} \le \tau \le 2$.
The treatment of nuclear liquid compressibility   allows us  to  reduce the tricritical endpoint density 
of the statistical multifragmentation model to one third of the normal nuclear density.  A specific attention is paid to of the fragment size distributions
in the region of a  negative surface tension  at  supercritical temperatures. 
\\

\end{abstract}

\maketitle

%
\section{Introduction}\label{secintro}

More than 30 years the statistical multifragmentation model (SMM) \cite{Bondorf:95} is playing
the leading role in studies of the nuclear multifragmentation reactions 
\cite{Gross:97,Moretto:97} which, probably, is one of the most spectacular phenomena that is available for  exploration  in nuclear  reactions  at  intermediate energies.
Additionally, the SMM greatly stimulated    the  studies  of phase transition (PT)  in finite systems \cite{Gross:97,Moretto:97,Randrup:04,Bugaev:07} and 
 investigation of  nonhomogeneous phases of strongly interacting matter in astrophysics \cite{Igor:2010a,Igor:2013b,Igor:AstroC} and heavy-ion collisions \cite{Mishustin:2006ka, Mishustin:1998eq,Mishustin:2008d}.
  
 A  simplified version of the SMM without the Coulomb and asymmetry terms was proposed in  \cite{simpleSMM:1, simpleSMM:1b}. Its analytical solution was  obtained in \cite{Bugaev:00,Bugaev:00b}, where an additional parameter, 
the Fisher exponent $\tau$, was introduced in the model.  The  value of  Fisher exponent extracted from nuclear  experiments \cite{ISIS, EOS:00}  turned out to be  $\tau \simeq 1.8 -1.9$, i.e. well below the 
prediction of the Fisher droplet model (FDM) $\tau_F \simeq 2.209 $ \cite{Fisher:67, Elliott:06}, but in a good correspondence 
with the  critical exponent  analysis   made in \cite{Reuter:01}  for   the simplified SMM.
 This  fact 
  initiated new attempts  to analyze the multifragmentation  data in order  to  extract  the values of  the exponent  $\tau$ and critical temperature 
\cite{Ogul:2002ka,Karnaukhov:03, Igor:2006xc}. These studies     gave a first  evidence  that 
the nuclear liquid-gas PT  has the tricritical endpoint rather than the critical one \cite{Reuter:01}.

The  successes achieved by  the SMM  in describing many sets of different  experimental data gave enough proofs of  its validity and this very fact 
prevented researches from asking two simple conceptual  questions: why does the SMM 
 perfectly work  in  describing the low density nuclear vapour that appears 
after the multifragmentation, and why the SMM  is unable to describe the high 
densities of nuclear matter achieved, for example, in heavy ion collisions. 
At first glance the answer on the second  question seems to be very trivial:
the SMM considers nucleons and  all nuclear fragments as incompressible  objects. 
This can be seen from the excluded volume of  the  $k$-nucleon fragment which is 
$V_k = \frac{ k}{\rho_0}$, where $\rho_0 \simeq 0.16 $ fm$^3$ denotes the normal nuclear density.  Thus, the limiting particle density appears in the SMM due to the Van der Waals like treatment of the short range repulsion between the fragments. However, in this case we really face a difficult problem to get the answer on the first of the above questions and to explain the reason why the SMM is so good in describing the experimental data at freeze-out density which is between 
$\frac{1}{12}$ and $\frac{1}{6}$ of  $\rho_0$.  It is well known that to describe 
the thermodynamics at low densities one has to use the virial expansion and account, at least,  for the second virial coefficients $b_{jk} = \frac{2}{3} \pi \left(R_j + R_k \right)^3$  between all pairs of  fragments of the hard core  radii $R_j$ and $R_k$ \cite{Stanley:71, Bugaev:RVDW1, Bugaev:RVDW2,Bugaev:RVDW3}.  The  real problem, however,  is that the SMM employs  not the second virial coefficients which  provide the description of  low density matter,
but  uses the proper volumes  of  the $k$-nucleon fragments  $V_k = \frac{4}{3} \pi R_k^3$, which, usually,
enter in  the high density limit \cite{Bugaev:07a,Bugaev:09a}!
Therefore, in order to understand why the SMM is  successful at low densities, first of all, 
we have to return to its basic assumptions  and find out,  how   the  virial 
coefficients appear in this model.   

The simplified version of the SMM \cite{simpleSMM:1,simpleSMM:1b} which is solved analytically  in  \cite{Bugaev:00,Bugaev:00b} is much more elaborate  than the FDM   \cite{Fisher:67} since, in contrast to the latter one,  the SMM  explicitly  contains   the nonzero proper sizes of  all fragments and, hence,  the liquid phase. However, in the standard SMM  the nuclear liquid is incompressible, that  is 
too rough approximation at higher temperatures. 
Moreover, the critical and tricritical endpoints of  the  simplified SMM
\cite{Bugaev:00, Reuter:01} appear  at the density of a liquid phase $\rho= \rho_l (T=0) = \rho_0 $, while  in ordinary substances 
the critical density is about one third of that one for  low temperature liquid phase  \cite{Stanley:71,Dillmann:91,LFK:94}. 
In present paper   we extend the SMM to account for the compressibility of nuclear liquid and 
show how the equation for the  surface tension coefficient  induced by interaction between the nuclear fragments  naturally appears from the virial expansion. Such an analysis  allows us to answer  the two conceptual questions formulated above. 
The developed approach has common features with Refs. \cite{Fisher:67, Dillmann:91,LFK:94}, but compared   to these works 
we  obtain  an  equation for the surface tension coefficient which is induced by interaction between the nuclear fragments. Therefore,  the developed model  can be considered  as a  further refinement of the ideas formulated in  \cite{Fisher:67, Dillmann:91,LFK:94}.  

The work is organized as follows.  In Section 2 we give a heuristic derivation of the equations for the  gaseous phase pressure  and for   the surface tension induced by the hard core repulsion. The conditions of the 1-st and 2-nd order PT   of  a liquid-gas type are formulated  in Section 3. The properties of the liquid phase pressure and the phase diagrams  for 
various  parameterizations  of the present model are discussed in Section 4. Section 5 is devoted to an  analysis of the fragment size distributions when going from subcritical to supercritical temperatures. 
The conclusions are formulated in Section 6.

%
%
\section{SMM with hard core repulsion}\label{secmodel}

In order to recapitulate the formal steps of  obtaining the Van der Waals EoS in the grand canonical 
ensemble, let us consider the one-component gas with the hard core repulsion.  The pressure of such a gas  with the temperature $T$ and chemical potential $\mu$  in the nonrelativistic approximation is given by 
\begin{eqnarray}\label{EqI}
p=T \, \phi (T) \exp\left[ \frac{\mu-a\, p}{T} \right] \,, \quad {\rm with} \quad \phi (T) = g \int \frac{d^3 \, \mathbf k}{(2\pi)^3}~\exp\left[  -\frac{\mathbf{k}^2}{2\, m\, T} \right] \,. 
\end{eqnarray}
Here  $a$ denotes the second virial coefficient,   $\phi (T)$  is a thermal density 
of particles having the mass $m$ (below this will be the nucleon mass $m = 940$ MeV) and the degeneracy factor $g$.
For  low  densities Eq. (\ref{EqI}) can be obtained from the virial expansion \cite{Stanley:71} at low densities 
\begin{eqnarray}
\label{EqII}
p \simeq  T \, 
\phi \, e^{\frac{\mu}{T}}\left(1-a\, \phi \, e^{\frac{\mu}{T}}\right) \,, 
\end{eqnarray}
in the following sequence of steps: first, one approximates the particle density as $\phi \, e^{\frac{\mu}{T}} \simeq \frac{p}{T} $, using the fact that at low densities such approximation is obeyed; second, now the obtained term is further approximated as  $1 - a\, \frac{p}{T} \simeq \exp\left[ - a\, \frac{p}{T}  \right] $. As a  result Eq. (\ref{EqI}) is reproduced. The final step is to extrapolate Eq. (\ref{EqI}) to high densities. In fact, the above result can be obtained from the traditional Van der Waals EoS without attraction 
either by its direct integration \cite{Bugaev:RVDW1} or 
by the maximum term method \cite{Bugaev:RVDW2}.

Let us apply the same steps  to the system of $N$-sorts  of  particles of  the hard core radii $R_k$, with $k = 1, 2, ..., N$. Then the virial expansion of the gas pressure  up to second order in particle density  is given by
\begin{eqnarray}\label{EqIII}
p   =T \sum_{k=1}^{N} \phi_{k}\, e^{\frac{{\mu}_{k}}{T}}\left(1- \sum_{n=1}^{N}\, a_{kn }\,  \phi_{n} \, e^{\frac{{\mu}_{n}}{T}} \right) \,,
\end{eqnarray}
where  $\phi _n(T) = g_n \int \frac{d^3 \, \mathbf{k}}{(2\pi)^3}~\exp\left[  -\frac{\mathbf{k}^2}{2\,m_n\, T} \right] $
denotes the thermal density of particles of the degeneracy $g_n$ and mass $m_n$, and 
$a_{kn}$ is  the second  virial coefficients of particles $k$ and $n$
\begin{eqnarray}\label{EqIV}
a_{kn}= \frac{2}{3} \pi \left(R_{k}+R_{n}\right)^{3}=\frac{2}{3}  \pi \left(R_{k}^{3}+3R_{k}^{2}R_{n}+3R_{k}R_{n}^{2}+R_{n}^{3} \right).
\end{eqnarray}
Of course, one can  straightforwardly repeat the same steps as before with the only modification that
for the multicomponent system  each sort of particles, say $n$,  generates its own pressure $p_n$ which 
should replace  in Eq. (\ref{EqIII}) the particle density as $ \phi_{n} \, e^{\frac{{\mu}_{n}}{T}} \simeq \frac{p_n}{T}$. Then one obtains the system of  equations for partial pressures  
\begin{eqnarray}\label{EqV}
p_k  =  T \,  \phi_{k}\, \exp \left[ \frac{{\mu}_{k}}{T} -  \sum_{n=1}^{N}\, a_{kn }\,  \frac{p_n}{T}  \right]  \,.
\end{eqnarray}
Such a system is known as the Lorentz-Berthelot mixture \cite{Bugaev:RVDW2} for which the total pressure is the sum of all partial ones $p = \sum_{k=1}^{N}\, p_k$.   It is clear that, in contrast to the one-component Van der Waals EoS, the inclusion of many-nucleon fragments accounts, at least partly, for  the attractive interaction between their constituents. 

The procedure of  the Van der Waals extrapolation is, however,  not unique and one can analyze  different  possibilities \cite{Bugaev:RVDW2,Bugaev:RVDW3}. Note that an order of  mathematical operations is important 
\cite{Bugaev:RVDW3}.  If, in contrast to above treatment,  one first  explicitly substitutes the second virial coefficients (\ref{EqIV}) into  expression for pressure  (\ref{EqIII}) first  and  regroups the powers of  radius $R_k$ of  a  
$k$-nucleon fragment, then one would find
\begin{eqnarray}
\label{EqVI}
p=T\sum_{k=1}^{N} \phi_{k}e^{\frac{{\mu}_{k}}{T}}\left[1-
\frac{4}{3}\pi R_{k}^{3}\cdot\sum_{n=1}^{N}\phi_{n}e^{\frac{{\mu}_{n}}{T}}-
4\pi R_{k}^{2}\cdot\frac{1}{2}\sum_{n=1}^{N}R_{n} \phi_{n}e^{\frac{{\mu}_{n}}{T}}-
2\pi R_{k}\cdot\sum_{n=1}^{N}R_{n}^{2} \phi_{n}e^{\frac{{\mu}_{n}}{T}}\right] \,. 
\end{eqnarray}
Now one can see that the right hand side (r.h.s.) of  Eq.  (\ref{EqVI}) contains the expansion 
in powers of radius $R_k$. If  one makes the Van der Waals extrapolation  of  pressure (\ref{EqVI})  in the same way, as we discussed earlier, then one can formally get the same structure of  the surface  free energy  which was suggested in  \cite{Dillmann:91}.  Thus,    Eq.  (\ref{EqVI})  contains the surface and curvature terms, which are, respectively,   proportional to $R_{k}^{2}$ and $R_{k}$ in the square brackets on its  r.h.s. 
However, in contrast to \cite{Dillmann:91,LFK:94} and their followers,  
the r.h.s. of  Eq.  (\ref{EqVI})  contains  the bulk term as well which is 
proportional to $R_{k}^{3}$  or to the $k$-nucleon  fragment  volume. Hence, all  fragments in  (\ref{EqVI}) are interacting with each other via a hard core repulsion. 
Now it is clear that, if, in addition,  one introduces the attraction via the spherical potential well of finite depth to model the proximity type  interaction between the nuclear fragments, then the only bulk term would not be modified, while the surface and curvature terms would get an additional  negative contributions which will be studied elsewhere.

Below we do not consider   the curvature term explicitly, but we   account   for  it  
implicitly by doubling 
 the   induced   surface free energy coefficient $ \Sigma$ and  introducing 
\begin{eqnarray}\label{EqVII}
p&=&T\sum_{k=1}^{N} \phi_{k}e^{\frac{{\mu}_{k}}{T}}\left[1-
\frac{4}{3}\pi R_{k}^{3}\cdot\frac{p}{T}-4\pi R_{k}^{2}\cdot \frac{\Sigma}{T} \right]
\nonumber \\
&\simeq&T\sum_{k=1}^{N} \phi_{k}
\exp\left[ \frac{{\mu}_{k}}{T} -\frac{4}{3}\pi R_{k}^{3}\cdot\frac{p}{T}-4\pi R_{k}^{2}\cdot \frac{\Sigma}{T} \right] \,. 
\end{eqnarray}

In order to guarantee a consistency   with the  derivation above,  we assume that 
 the induced  surface free energy  coefficient  
$\Sigma$ obeys  the following equation
\begin{eqnarray}
\label{EqVIII}
\alpha \, \frac{\Sigma}{T} & \simeq  & \alpha\,  \sum_{k=1}^N  R_k \, \phi_{k} \, \exp\left[ \frac{{\mu}_{k}}{T} -\frac{4}{3}\pi R_{k}^{3}\cdot\frac{p}{T}-4\pi R_{k}^{2}\cdot \alpha\, \frac{\Sigma}{T} \right] \,. 
\end{eqnarray}
Here the constant $\alpha > 0$  is introduced due to the freedom of the Van der Waals extrapolation to high densities.
In this way the present model  accounts for higher order corrections compared  to the low density virial expansion.  As  will be shown later,  such a correction plays an important role at the vicinity of the tricritical endpoint. 
Also it is important to note that in (\ref{EqVIII}) the  induced  surface free energy  coefficient
is  extrapolated to high densities using the  same ensemble of  one-particle distribution functions  $\phi_k$ for $k$-nucleon fragments   as the  one employed in (\ref{EqVII}).   Now it is clear that at low particle densities Eqs. (\ref{EqVII}) and  (\ref{EqVIII}) correctly account for the virial expansion up to the second  order   while the deviation appears at the third virial coefficient.

To connect   the above formula   for pressure (\ref{EqVII})  with   the gaseous phase  pressure of  the SMM, we  parameterize the one-particle thermal  densities of  all  $k$-nucleon fragments as  
\begin{eqnarray}
\label{EqIX}
 \phi_{1}  = z_1 \left[ \frac{m T}{2\pi}\right]^{\frac{3}{2}}  \exp \left[   -  \frac{\sigma_0 (T)}{T}   \right]\, , \quad 
  \phi_{k \ge 2}  = g  \left[ \frac{m T}{2\pi}\right]^{\frac{3}{2}}  \frac{1}{k^\tau} 
  \exp \left[  \frac{\left( k\, p_L V_1 - \mu_k \right)}{T} -  \frac{\sigma_0 (T)}{T}k^ \varkappa   \right] \,, 
\end{eqnarray}
where $z_1 = 4$ is the degeneracy factor of nucleons, while the degeneracy factor for other fragments $g$  is, for simplicity, chosen to be 1. 
In the final step of  our  heuristic derivation of  the system of Eqs.  (\ref{EqVII}) and  (\ref{EqVIII}) 
we  presented the $k$-nucleon ($k>1$) volume as $V_k = \frac{4}{3}\pi R_k^3 = V_1 k$, where $V_1 = \frac{1}{\rho_0}$. 
 The binding energy per nucleon of  such a  fragment  is expressed via   the free energy of liquid drop $- k V_1 p_L (T, \mu)$ of same size and  the corresponding   chemical potential $\mu_k$.  
In the standard SMM  the  pressure of liquid phase is $p_L (T, \mu) = \frac{\mu + W(T)}{V_1}$ \cite{Bugaev:00} and the $k$-nucleon fragment chemical potential  $\mu_k = k \, \mu$ is expressed via the  chemical potential of nucleon $\mu$. Obviously, for the standard SMM one finds the usual result  for binding free  energy $ k\, p_L V_1 - \mu_k  = k \, W (T)$.  
In Eq. (\ref{EqIX}) the power  $\tau$ is the Fisher topological exponent \cite{Fisher:67}, which was introduced into  the SMM in  \cite{Bugaev:00}.  In actual simulation we use the value $\tau = 1.9$ which is motivated  by the experimental data 
\cite{ISIS, EOS:00} (see below).

It is necessary to remind that  in the standard SMM the temperature dependent binding energy is  $W (T)=W_0 + 
W_{Fm} (T)$. Here $W_0=16 $ MeV is the bulk binding energy per nucleon at vanishing temperature, while the term $W_{Fm} (T) \equiv \frac{T^2}{\varepsilon_0}$ is the contribution of the excited states taken in the Fermi-gas approximation  ($\varepsilon_0=16$ MeV).
The r.h.s. of Eq.  (\ref{EqIX}) contains  the  liquid phase pressure $p_L(T,\mu)$ of  general form. Such   a generalization of the SMM allows one  to consider more complicated $\mu$ and $T$ dependencies of  the  liquid phase  pressure than the one used in
the original SMM. 
It is important that the resulting model automatically obeys the L. van Hove axioms of  statistical mechanics \cite{Hove}, if  in the liquid phase  the liquid pressure obeys these axioms (see a discussion below).
Due to this  property  the present model does not lead to an appearance of non-monotonic isotherms in the density-pressure plane  which are  typical for mean-field  models of equation of state.

In Eq.   (\ref{EqIX})  we have  introduced  the eigen part of  the   temperature-dependent surface tension $ \sigma_0(T)$ for each fragment including the  nucleons.  This is a slight change compared to the standard SMM formulation \cite{Bondorf:95, Bugaev:00} which, however, does not strongly affect   properties of the phase diagram, but somewhat  simplifies the analysis of the  model. 
The parameter $\varkappa$ is chosen to match that one of the SMM,  i.e.  $\varkappa =  \frac{2}{3}$, but in the next section we   analyze the properties of the PT diagram for  a wider range of  its values, namely for  $ \frac{2}{3} \le   \varkappa < 1$ .   

Collecting all terms together, we rewrite Eqs. (\ref{EqVII}) and (\ref{EqVIII}) as the following system  
\begin{eqnarray}
\label{EqX}
&&\hspace*{-0.5cm}\frac{p}{T} =   \left[ \frac{m T}{2\pi}\right]^{\frac{3}{2}}  \sum_{k=1}^{N}  \frac{b_k(T)}{k^\tau}
\exp\left[ \frac{ \left( p_L - p \right) V_1\, }{T}k - \frac{\left(\sigma_1 + \sigma_0\right) }{T}k^\varkappa  
\right] \,,  \\
\label{EqXI}
&&\hspace*{-0.5cm}\frac{\alpha \,\sigma_1}{3 \,T\, V_1}  =  \alpha \,   \left[ \frac{m T}{2\pi}\right]^{\frac{3}{2}}
 \sum_{k=1}^{N}  \frac{b_k(T)}{k^{\tau -\frac{1}{3}} }
\exp\left[ \frac{ \left( p_L - p \right) V_1\, }{T}k - \frac{\left(\alpha \, \sigma_1 + \sigma_0\right) }{T}  k^\varkappa  
\right] \,,  \quad \quad 
\end{eqnarray}
for an infinite number of  the sorts of nuclear fragments  $N \rightarrow \infty$.
Here the temperature dependent degeneracy is defined as $b_1 (T) \equiv 4\,  \exp\left[ \frac{ -W(T)}{T} \right]$ and 
$b_{k>1} (T) = 1$.  
In this way we account for the 
fact that, compared to larger fragments,  the nucleons have  no binding energy. 
One should bear in mind  that for an infinite number of  the sorts of nuclear fragments  $N$ a  redefinition of any finite number of   internal partitions   of  $k$-nucleon  fragments, i.e. $b_{k} (T)$, 
does not affect the divergency or convergency  of  the pressure  (\ref{EqX})  and the induced surface tension (\ref{EqXI}).

Choosing  some temperature dependent parameterization for  the eigen surface tension coefficient $\sigma_0(T)$ for the fragments, we obtain the  closed system of equations to determine the pressure $p$  and 
the total surface tension coefficient $\sigma_0 + \sigma_1$ for a given value of parameter $\alpha$.
The pressure in this  system is closely resembling the expression suggested in  the  simplified SMM to study 
the nuclear liquid-gas PT in thermodynamic limit \cite{Bugaev:00, Reuter:01, simpleSMM:1, simpleSMM:1b}.

The  above consideration  clearly demonstrates that the liquid drop parameterization of the internal free energy employed 
 in the original SMM  
\cite{Bondorf:95} is consistent with the Van der Waals extrapolation where  
the  proper volume
of fragments is used  for the bulk part,  while the rest   of the hard core repulsion is  accounted  by 
the corresponding  choice of  the surface free energy. Similarly, the  attractive  interaction between the fragments  is partially recorded  in the  surface  free energy,
while another part of attractive interaction  is  stored in the very fact that   fragments   with a certain  binding energy
are formed. 
By construction the EoS  defined by (\ref{EqX}) and  (\ref{EqXI})    automatically reproduces  the low density expansion. 
In addition, the suggested EoS contains the bulk part of free energy, which is applicable   at high densities, and, consequently,   such EoS  may correctly reproduce  properties of  the liquid phase 
that  is impossible in  the model of non-interacting clusters \cite{Fisher:67,Dillmann:91,LFK:94}. 

\section{Conditions for  the tricritical point existence}

In order to study the nuclear liquid-gas PT~  Eqs.  (\ref{EqX}) and   (\ref{EqXI}) should be supplemented  by the EoS  of  the liquid phase pressure $p_L(T,\mu)$ to be substituted 
into an exact analytical expression  of the simplified SMM found in \cite{Bugaev:00, Reuter:01}.  However, here we would like to follow  a more traditional way for  analyzing  the necessary and sufficient  conditions for  the nuclear liquid-gas PT. For this purpose  we
rewrite  (\ref{EqX}) and   (\ref{EqXI}) in terms of dimensionless  variables $\xi_1$, $\xi_2$ and $\tilde \xi_2$
\begin{eqnarray}
\label{EqXII}
\xi_1&=& -\xi_L+V_{1}I_\tau(\xi_1,\xi_2) \,, \\
\label{EqXIII}
\xi_2&=& \sigma_0+ 3\, V_1 I_{\tau-\frac{1}{3}}(\xi_1,\tilde \xi_2)\, , \\
\label{EqXIV}
I_{\tau-q}(\xi_1,\xi_2) & \equiv & \left[ \frac{m T}{2\pi}\right]^{\frac{3}{2}}\sum_{k=1}^{N}  \frac{ b_k(T) }{k^{\tau-q}} \exp\left[-\xi_1 k-\xi_2 k^ \varkappa \right] \, .
\end{eqnarray}
Here the dimensionless variables $\xi_1$,  $\xi_2$ and $\tilde \xi_2$ are  defined in terms of the bulk and surface components of  the  fragment's  free energy taken per one nucleon  as
\begin{eqnarray}
\label{EqXV}
\xi_1&\equiv&   \frac{p V_1}{T} - \xi_L \,, \quad  \xi_L \equiv  \frac{p_L  V_1}{T} \,, \\
\label{EqXVI}
\xi_2&\equiv & \frac{\sigma_0+ \sigma_1}{T} \,,   \hspace*{7.50mm} \tilde \xi_2 \equiv  \frac{\sigma_0+ \alpha\, \sigma_1}{T}   \, . 
\end{eqnarray}
The variable $\xi_1$ is convenient for an  analytical  manipulations since in terms of this variable  the 
  gas and liquid coexistence  condition    $p  (T, \mu) = p_L (T, \mu) $  reads as  $\xi_1 (T, \mu) =0$.
Its solution  $\mu = \mu_c (T)$ defines the phase diagram in the $T-\mu$ plane. Similarly, the variable $\xi_2$ which according to (\ref{EqXIII}) describes    the  surface free energy coefficient  as $T \xi_2$  is  convenient to  detect the critical  (or tricritical) endpoint of the phase diagram. Indeed, below we explicitly demonstrate, that  for $\xi_2 (T, \mu_c (T)) > 0$
the PT is of the 1-st order, whereas for  $\xi_2 (T, \mu_c (T)) = 0$  the PT is of the 2-nd or higher order.  
The auxiliary variable $\tilde \xi_2 \equiv \frac{\alpha\,  \xi_{2}+(1-\alpha)\sigma_{0}}{T}$ is convenient  both for analytical and   numerical  evaluations since  instead of Eq. (\ref{EqXIII}) it is more appropriate to  solve the  equation 
\begin{eqnarray}
\label{EqXVII}
\tilde{\xi}_{2}=\sigma_{0}+3\, \alpha \,V_{1}{J}_{\tau-\frac{1}{3}}(\xi_1,\xi_2) \,, 
\end{eqnarray}
where we introduced the following notations ${J}_{\tau-q}(\xi_1,\xi_2) \equiv {I}_{\tau-q}(\xi_1,\tilde{\xi}_2)$.

Let us now show that the  equation  $\xi_1 (T, \mu)  =0$ corresponds to a PT.  Assuming that its solution $\mu = \mu_c (T)$ exists we study the  necessary conditions for a PT occurrence. For this purpose it  is sufficient to study the properties of the partial  $\mu$-derivatives of  the variables $\xi_1$ and $\tilde \xi_2$. The  analysis of  the partial $T$-derivatives  leads to the same results although the obtained equations are more involved. From Eqs. (\ref{EqXII}) and  (\ref{EqXVII}) one  finds  
\begin{eqnarray}
\label{EqXVIII}
\frac{\partial\xi_{1}}{\partial\mu}
&=&
-\frac{\left(1+3V_{1}\alpha{J}_{\tau-\varkappa-\frac{1}{3}} \right) \, \frac{\partial \xi_{L}}{\partial\mu}}{\left(1+V_{1}I_{\tau-1}\right) \left(1+3V_{1}\alpha{J}_{\tau-\varkappa-\frac{1}{3}} \right)-3V_{1}^{2}\alpha{J}_{\tau-\frac{4}{3}}I_{\tau-\varkappa}},\\
\label{EqXIX}
\frac{\partial\tilde{\xi}_{2}}{\partial\mu}
&=&-\frac{3V_{1}\alpha{J}_{\tau-\frac{4}{3}}\, \frac{\partial \xi_{L}}{\partial\mu} }{(1+V_{1}I_{\tau-1}) \left(1+3V_{1}\alpha{J}_{\tau-\varkappa-\frac{1}{3}} \right)-3V_{1}^{2}\alpha{J}_{\tau-\frac{4}{3}}I_{\tau-\varkappa}}.
\end{eqnarray}
According to the standard definition of statistical mechanics, if  the 1-st derivatives of  the pressure of  two phases differ from each other  at their  coexistence  curve, then this  is a 1-st order PT.  If, however, at the two-phase coexistence curve   the  1-st derivatives of  the pressure of  two phases coincide   then a PT is of the 2-nd or higher order.  Eq. (\ref{EqXVIII}) provides one with the 
difference of the particle number  density of  gaseous $\rho_g$ and liquid $\rho_L$  phases, since  $\frac{\partial {\xi}_{L}}{\partial\mu} \equiv  \rho_L V_1/T$ and  $\frac{\partial {\xi}_{1}}{\partial\mu} \equiv  (\rho_g - \rho_L) V_1/T$.

To simplify the analysis we assume that in  Eq. (\ref{EqXVIII})  $\alpha > 1$ and will comment in appropriate places what occurs for  the case   $\alpha \le 1$.  Then all the sums  ${I}_{\tau-q}(0, \xi_2)$ and  ${J}_{\tau-q}(0, \xi_2)$ are finite for 
$\xi_2 > 0$.  The validity of this statement for the sums  ${I}_{\tau-q}(0, \xi_2)$ follows from a direct inspection of Eq. (\ref{EqXIV}), while for the sums  ${J}_{\tau-q}(0, \xi_2)$ it follows from the fact that for $\alpha > 1$ the induced surface tension term $\sigma_1$ defined by  (\ref{EqXI}) is always positive and, hence, $\tilde \xi_2 > \xi_2$.  Therefore,   for  $\xi_1 = 0$ and $\xi_2 > 0$  the r.h.s.  of  Eq.  (\ref{EqXVIII}) does not vanish, i.e. $\frac{\partial\xi_{1}}{\partial\mu} \neq 0$, and, hence,   in this case 
we are dealing with  the 1-st order PT
and the present model is similar to the simplified SMM  \cite{Bugaev:00,Reuter:01}.

A  1-st order PT occurs also in  the case of  intersecting  curves $\xi_1 = 0$    and   $\xi_2 = 0$ at  $T=T_{cep}$  for $\tau > 2$.  This can be seen form Eqs. (\ref{EqXVIII}) and (\ref{EqXIX}), since for $\tau > 2$   all sums ${I}_{\tau-1}(0, 0)$ and  ${I}_{\tau-\varkappa}(0, 0)$ on the r.h.s. of these equations are finite. Therefore, at $T = T_{cep}$ 
this case exhibits  not a critical point, but a critical line due to different particle densities of the gaseous and liquid phases.
Such phase diagrams are  known from   the literature  \cite{Fisher:70, Bugaev:07,Bugaev:07a}, although there are no such empirical examples.


A different situation arises  in the case $\tau \le 2$, if  on  the curve $\xi_1 = 0$    one also  has a point where  $\xi_2 = 0$.  Then the sums  ${J}_{\tau-q}(0, \xi_2)$ are finite since $\tilde \xi_2 > \xi_2$, but  the sums ${I}_{\tau-1}(0, 0)$ and  ${I}_{\tau-\varkappa}(0, 0)$ diverge.  Thus, for  $\tau \le 2$  the sum ${I}_{\tau-1}(0, 0) \rightarrow \infty$ and in this case $\frac{\partial\xi_{1}}{\partial\mu} = 0$ and, hence,   the model  exhibits a PT of the 2-nd  or higher order. Note that by the adopted  assumptions 
$\varkappa < 1$, the sum  ${I}_{\tau-1}(0, 0)$ is the most divergent one in  Eq.  (\ref{EqXVIII}). 
After evaluating  the second derivative of $\xi_1$  and taking the limit $\xi_2 \rightarrow +0$, one obtains 
\begin{eqnarray}
\label{EqXX}
\frac{\partial^2 \xi_{1}}{\partial\mu^2}  
& \rightarrow & - \frac{\partial  {\xi}_{L}}{\partial\mu}\cdot \frac{1}{V_{1}I^{2}_{\tau-1}} \left[  \frac{\partial\xi_{1}}{\partial\mu} I_{\tau-2}  +    \frac{1}{\alpha} \frac{\partial \tilde \xi_{2}}{\partial\mu} I_{\tau-1-\varkappa} \right] 
 \rightarrow  -\left[ \frac{\partial  {\xi}_{L}}{\partial\mu}\right]^{2}\cdot \frac{I_{\tau-2}(0,0)}{V_{1}^{2}[ I_{\tau-1}(0,0)]^3} \, ,
\end{eqnarray}
where in the last  step of evaluation we took   into account   that  the sum $I_{\tau-2}(0,0)$ is more divergent than the sum $I_{\tau-1-\varkappa}$ and that according to Eq.  (\ref{EqXIX}) 
\begin{eqnarray}
\label{EqXXI}
\frac{\partial\tilde{\xi}_{2}}{\partial\mu}  & \rightarrow & \frac{\partial\xi_{1}}{\partial\mu}\cdot \frac{3V_{1}\alpha{J}_{\tau-\frac{4}{3}}}{1+3V_{1}\alpha{J}_{\tau-\varkappa-\frac{1}{3}}} \, .
\end{eqnarray}
To complete the evaluation of    Eq. (\ref{EqXX}) we need to analyze the behavior  of the sums $I_{\tau-q}(0,\xi_2)$  in the limit $\xi_2 \rightarrow 0$.  This can be done  either rigorously using the integral representations of  these sums developed in \cite{Reuter:01}
or  employing a more straightforward way used below.  Consider first the large, but finite number of nuclear fragment sorts   $N$  in all sums (\ref{EqXIV}). Then each sum   is finite for any finite value of parameters and, hence,  one
 can find the limit $\xi_1 = 0$ and  $\xi_2 = 0$ without any trouble and calculate the sum as an integral 
\begin{eqnarray}\label{EqXXII}
I_{\tau-q}(0, 0) & = & \lim_{N\rightarrow \infty} \left[ \frac{m T}{2\pi}\right]^{\frac{3}{2}}\left[  b_1(T) +  \sum_{k=2}^{N}  \frac{ 1}{k^{\tau-q}}    \right]   =  \left[ \frac{m T}{2\pi}\right]^{\frac{3}{2}} \, b_1(T) 
  +      \lim_{N\rightarrow \infty} \left[ \frac{m T}{2\pi}\right]^{\frac{3}{2}}  \int\limits_2^N d x \, x^{q-\tau} \nonumber \\
&   =  &
\left[ \frac{m T}{2\pi}\right]^{\frac{3}{2}} \, b_1(T)   + \left[ \frac{m T}{2\pi}\right]^{\frac{3}{2}} \cdot \lim_{N\rightarrow \infty}
\left\{
\begin{array}{ll}
\frac{1}{\tau - q -1} 2^{q+1 -\tau } \, ,  &  {\rm for} \quad   q < \tau -1 \,,   \\
 & \\
 \ln \left[  \frac{N}{2} \right]  \, ,   &    {\rm for} \quad   q = \tau -1 \,,  \\
  &      \\
  \frac{1}{q+1 -\tau } N^{q+1 -\tau } \, ,  &  {\rm for} \quad   q > \tau -1 \,.
\end{array}
\right.
\end{eqnarray}
Applying these results to the case $\tau \le 2$, one finds  that in the limit $N\rightarrow \infty$  both sums $I_{\tau-2}(0,0) \sim N^{3-\tau}$  and 
$I_{\tau-1}(0,0) \sim N^{2-\tau}$ diverge, but the second derivative $\frac{\partial^2 \xi_{1}}{\partial\mu^2} \sim  N^{2 \tau -3}$ in  (\ref{EqXX}) vanishes only   for $\tau <  \frac{3}{2}$. Therefore, in this model   the 2-nd order PT exists  for $\xi_1 = 0$,  $\xi_2 = 0$ and   $\frac{3}{2} \le  \tau \le 2$, whereas for  $\tau < \frac{3}{2}$ there is a PT of  the 3-rd or higher order.  Similarly, studying the  higher  order $\mu$-derivatives of  functions  $\xi_1$ for  intersecting 
(or matching) curves   $\xi_1 = 0$ and  $\xi_2 = 0$, one can get  the following expression 
\begin{eqnarray}
\label{EqXXIII}
\frac{\partial^{n}\xi_{1}}{\partial\mu^n}\sim-\left[ \frac{\partial \xi_{L}}{\partial\mu}\right]^n \cdot \frac{[I_{\tau-2}(0,0)]^{n-1}}{V_{1}^{n}[ I_{\tau-1}(0,0)]^{2n-1}} \, ,
\end{eqnarray}
for the $n$-th derivative $(n=3, 4, 5, \dots)$. 
In  this case there exist the $n$-th order PT for $\frac{n+1}{n} \le \tau < \frac{n}{n-1}$. The last result can be applied for $n=2$ as well, although   $\tau=2$  is a special case of the 2-nd order PT that is not described by this formula. Thus, the minimal value of the Fisher index is $\tau =1$ and in this case one  may expect  a Kosterlitz-Thouless PT of infinite order. 
Note that this situation  is different from the simplified SMM analytical solution \cite{Bugaev:00,Reuter:01} for which the critical endpoint exists for any $\tau < 1$, but it is similar to the gas of bags with surface tension model  which has   a tricritical endpoint \cite{Bugaev:07a}. 

Now it is necessary to mention that for $\alpha =  1$ the whole  previous consideration should be modified, since in this case  
for  $\xi_1=0$ and  $\xi_2= \tilde \xi_2 \rightarrow +0$
the  leading term in Eqs. (\ref{EqXVIII}) and  (\ref{EqXIX}) is not $I_{\tau-1} (0,\xi_2 )$, but the product
${J}_{\tau-\frac{4}{3}} (0,\xi_2) \, I_{\tau-\varkappa}( 0,\xi_2 )$. Our numerical analysis showed that although 
in this case there exists a tricritical endpoint  for  $\xi_1=0$ and  $\xi_2 = 0$ in its vicinity the denominator of expressions 
(\ref{EqXVIII}) and  (\ref{EqXIX}) always  changes a  sign leading to an unphysical situation, when the negative values of particle density of a  gaseous phase coexist with a  liquid phase of  positive particle density.  For  $\alpha <  1$  the model does not have the (tri)critical endpoint, since from  $\xi_1=0$ and  $\xi_2 = 0$ it follows that  $\tilde \xi_2 <  0$ due 
to positive values of $\sigma_1$ in (\ref{EqXVI}). However, for  $\tilde \xi_2 <  0$ and  $\xi_1=0$  the 
r.h.s.  of 
(\ref{EqXVII}) diverges to $+\infty$, whereas its left hand side is negative.  Consequently, such solutions do not exist. 
Therefore, in this work  we consider only the case $\alpha > 1$ which has no such  defects. 
In fact, the values  $\alpha > 1$ show that  the third and higher order virial coefficients 
are less important for the induces surface tension $\sigma_1$  at the tricritical endpoint  vicinity and, hence,  compared to the second virial coefficients their contribution to   $\sigma_1$ is  suppressed 
due to  the inequality   $\alpha > 1$.

It is easy to show that for  any finite value of  the  function $\xi_L$ there are no solutions with  $\xi_1 <  0$. Let us assume that such a solution 
of the system  (\ref{EqXII})-(\ref{EqXIV}) exists. Then for any value of $\tau$ and $\xi_2$ the sum $I_\tau(\xi_1,\xi_2)$ diverges to $+\infty$ and, hence, from  (\ref{EqXII}) one gets that $\xi_1 \rightarrow +\infty$, i.e. we arrive at  a contradiction.  However, the present model  contains   the solutions with negative surface tension coefficients, i.e.  for 
$\xi_2 <  0$. It is clear that for  $\xi_2 <  0$ there is no PT, since in this case the solution of   the system  (\ref{EqXII})-(\ref{EqXIV}) exists only  for $\xi_1 > 0$.  Compared to  the simplified SMM this is entirely new case, which at  first glance may look unphysical. However,  as it is argued  in \cite{Bugaev:07a,Bugaev_09} there is nothing wrong or unphysical with the negative values of surface tension coefficient, since in the grand canonical ensemble the quantity $\xi_2 \, 
k^\varkappa$ is the surface  free energy $f_{surf}=e_{surf}-Ts_{surf}$ of the nuclear fragment  of mean volume $k V_1$, were $e_{surf}$ and $s_{surf}$ are the surface energy and entropy. Therefore, $\xi_2 <0$ means  that the surface entropy contribution simply exceeds the surface energy part, i.e. $Ts_{surf} > e_{surf}$ and then $f_{surf}<0$.
It can be shown on the basis of exactly solvable model of surface deformations  \cite{Bugaev:04a, Bugaev:07b} that  negative values of the surface free energy  appear as  a consequence  of very large  number of non-spherical configurations at high temperatures. 

We would like to point out that   negative values of  the surface tension coefficient  may  provide us a physical  reason preventing the condensation of small droplets into a liquid phase (an infinite droplet) at supercritical temperatures, and  naturally explain  the existence of a cross-over transition in ordinary liquids \cite{Bugaev:Nucleation11}  as well as  in QGP   \cite{Bugaev:07a, Bugaev_09, Bugaev:12b,Bugaev:2012c}.  This may  lead
 to an appearance of  surfaces with the fractal dimension.

\section{Phase diagram of the nuclear liquid-gas PT}

In the original SMM    the  pressure of nuclear liquid is expressed as  $p_L=\frac{\mu+W(T)}{V_{1}}$. Such a parameterization 
corresponds to an incompressible liquid, since the isothermal compressibility  $K_{T}\equiv\frac{1}{\rho}\frac{d\rho}{dp}\mid_{T}$ is  zero in this case.  
Obviously, this fact is in contradiction with the existing experimental information on the compressibility  modulus. 
Here we propose another  liquid pressure parametrization which    provides 
a more realistic description. Since  
the necessary condition for a non-zero compressibility is    $\frac{\partial^2 p_L}{\partial\mu^2}\neq0$,    we assume that liquid phase pressure has a form \cite{Bugaev:Nucleation11}
\begin{eqnarray}
\label{EqXXIV}
p_L=\frac{W_{Fm}(T)+ \mu + W_0 +a_{\nu}\left[ \mu  + W_0 \right]^{\nu}}{V_1}\, ,  \quad {\rm for} \quad \nu = 2, 3, 4,
\end{eqnarray}
where $a_\nu$ is a constant which should be found from a normalization condition (more details can be found  in Appendix A).
Of course, one could use more complicated parameterizations of the liquid phase pressure, but below we show that
even such a simple  modification leads  to  a rather rich structure of  the nuclear   phase diagram. 

Using the definition of particle density  for the liquid phase,  $\rho_L=\frac{\partial p_L}{\partial\mu}$, one gets 
\begin{eqnarray}
\label{EqXXV}
\rho_{L}=\frac{1+a_{\nu}\nu \tilde \mu^{\nu-1}}{V_{1}} \,, 
\end{eqnarray}
where an effective chemical potential is denoted as $\tilde\mu \equiv \mu  + W_0$.
One can identically rewrite Eq.  (\ref{EqXXIV}) in a more familiar way
\begin{eqnarray}
\label{EqXXVb}
p_L= \rho_0 \left[ W_{Fm}(T)+ \left(  \frac{\rho_L - \rho_0}{ \nu\, a_{\nu}\rho_0 }   \right) ^\frac{1}{\nu-1} \left[ 1 +  \frac{\rho_L - \rho_0}{ \nu\, \rho_0 } \right]  \right] \, ,  
\end{eqnarray}
which for $\nu = 2$ is a second order polynomial in $\rho_L$, while for other  integer values of  the parameter $\nu$ 
Eq.    (\ref{EqXXVb}) looks more sophisticated.

An important practical purpose to employ the liquid gas pressure (\ref{EqXXIV}), (\ref{EqXXVb})  is to reduce the standard SMM critical density  $\rho_{cep}$
from the normal nuclear density $\rho_0 = 1/V_1 \simeq 0.16 $ fm$^3$ to the value $\rho_{cep} = \rho_0/3$ which is typical for the liquid-gas PTs \cite{Stanley:71}. The condition $\rho_{cep} = \rho_0/3$ is an additional constraint on the developed EoS. 
Using  (\ref{EqXXV}) and
requiring that at the critical endpoint with the coordinates  $(T_{cep}; \mu_{cep})$  the condition  $\rho_{L} (\mu_{cep}) = \rho_{cep}= \rho_0/3$, one finds  the normalization constant $a_\nu$ 
\begin{eqnarray}
\label{EqXXVI}
a_{\nu}=  \left[ \frac{\rho_{cep}}{\rho_0} - 1\right] \frac{ \tilde \mu_{cep}^{1-\nu}}{ \nu} =-\frac{2 \, \tilde \mu_{cep}^{1-\nu}}{3\, \nu} \,.
\end{eqnarray}
Since at the critical endpoint  $\xi_1 = \xi_2 =0$,    Eq. (\ref{EqXII}) can be written as 
\begin{equation}\label{EqXXVII}
a_{\nu}\tilde \mu^{\nu}+\tilde \mu+W(T_{cep}) - T_{cep} V_1 I_\tau(0,0)=0 \,.
\end{equation}
Thus, the chemical  potential at critical endpoint $\mu_{cep}$ can be found from Eq. (\ref{EqXXVII}) directly. 
Indeed, 
substituting the standard value of the critical temperature  $T_{cep} = 18$ MeV into (\ref{EqXXVII}) for an unknown  $\tilde \mu = \tilde \mu_{cep}$ and accounting for the normalization condition (\ref{EqXXVI}), one finds 
\begin{equation}\label{EqXXVIII}
\tilde \mu_{cep} \left[1 +  \frac{1}{ \nu}\left( \frac{\rho_{cep}}{\rho_0} - 1\right) \right] =  T_{cep} V_1 I_\tau(0,0) - W(T_{cep})\,.
\end{equation}
 Note that  Eq. (\ref{EqXXVIII}) has a single solution for  $\tilde \mu = \tilde \mu_{cep}$!
Finding the $\mu_{cep}$ values for different $\nu$,  one can determine the normalization constant $a_\nu$ from 
Eq. (\ref{EqXXVI}). Their values are given in Table 1.  
 However, if  one substitutes  now the obtained value of  $a_\nu$ into  Eq. (\ref{EqXXVII}) and solves  
it for $\tilde \mu$, then one obtains several solutions for $\mu_{cep}$!  For instance, for $\nu=2$ one can easily check that there are two solutions $\tilde \mu_{cep1} \simeq -26.5$ MeV and $\tilde \mu_{cep2} \simeq - 54$ MeV.   
 Note that an additional solution $\tilde \mu_{cep2}$ is unphysical since it corresponds to the negative particle density of the liquid.

\begin{table}[!]
\caption{\label{tab:bolts} The model parameters  providing the condition $\rho_{cep} = \rho_0/3$.}
\hspace*{0.3cm}
\begin{center}
\begin{tabular}{|c|c|c|c|}
\hline
model parameter &  \multicolumn{3}{c|}{value} \\
\hline
$\tau$ &  1.9 & 1.9 & 1.9 \\
\hline
$\nu$ &  2 & 3 & 4 \\
\hline 
$a_\nu$  (MeV$^{1-\nu}$) & 1.261 $ \cdot 10^{-2}$ & -4.414 $\cdot  10^{-4}$  & 1.803 $\cdot  10^{-5}$   \\
\hline   
\end{tabular}
\end{center}
\end{table}

\begin{figure}[ht]
\centering
\includegraphics[width=16cm, height=10cm]{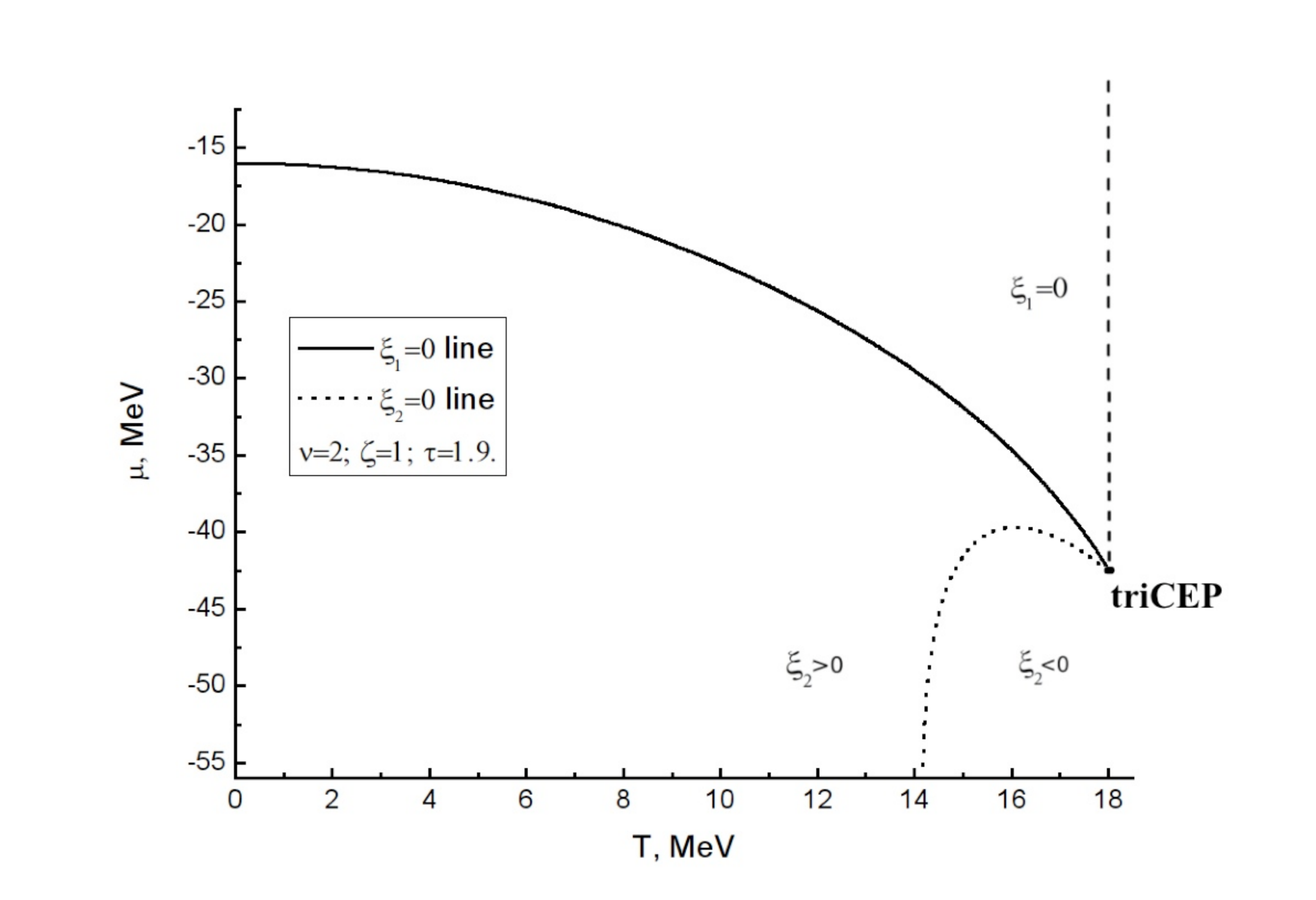}
\caption{Phase diagram in $T-\mu$ plane for the case $\tau$=1.9, $\nu=2$. At the critical temperature  $T_{cep}$=18 MeV there is   a triCEP. 
The solid  curve shows  a 1-st  order PT, the  long dashed one  shows  a  2-nd order PT, while the short dashed curve indicates the nil line of the surface tension coefficient.}
\label{Fig1}
\end{figure}

\begin{figure}[ht]
\centering
\includegraphics[width=16cm, height=10cm]{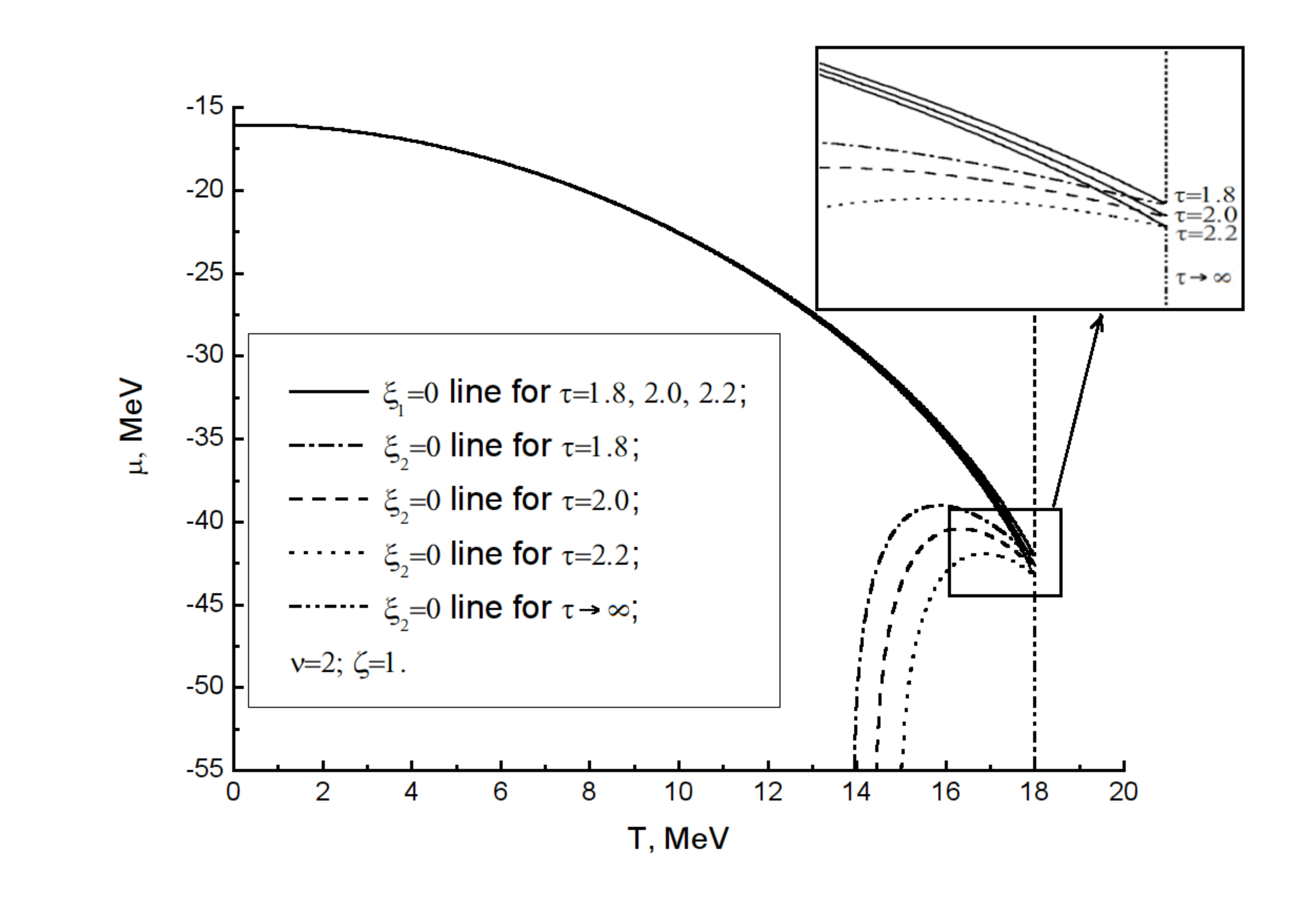}
\caption{Phase diagram in $T-\mu$ plane for $\nu=2$ is shown for several values of the Fisher topological parameter $\tau$. The line of the 2-nd order PT (vertical short dashed line) is shown   for $\tau =1.8$ only, since other lines   are hardly  distinguishable  from each other. Also the nil surface tension line of the limiting case $\tau \rightarrow \infty$  
is shown for a comparison by the  vertical dashed-double-dotted line.}
\label{Fig1b}
\end{figure}

\begin{figure}[ht]
\centering
\includegraphics[width=14cm, height=10cm]{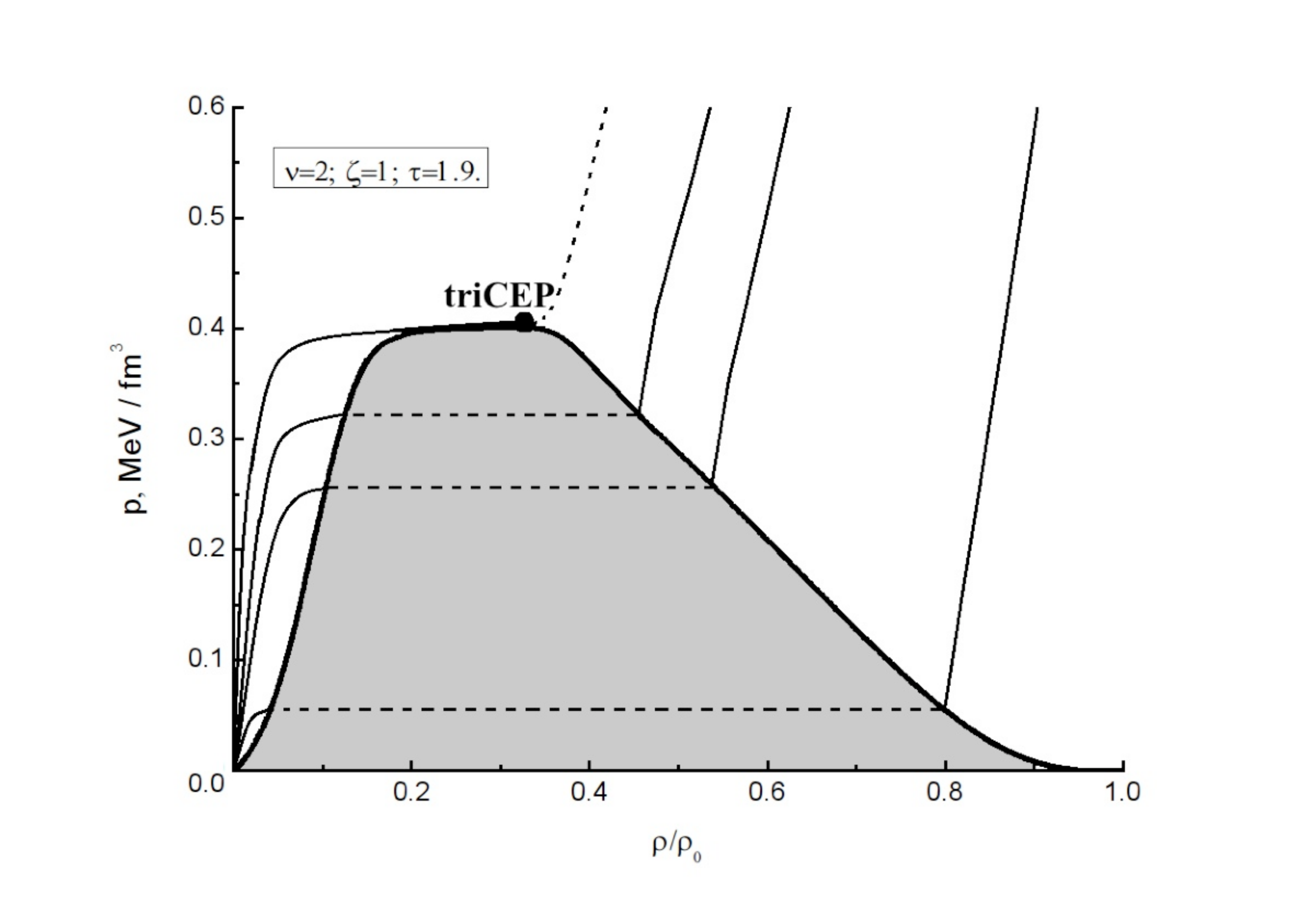}
\caption{Phase diagram in $\rho-p$ plane for $\nu=2$ and $\tau =1.9$.
The grey area corresponds to a mixed phase of the 1-st order PT.
 The isotherms are shown for T=11,16,17,18 MeV from bottom to top.   For the density  $\rho/\rho_0  \ge 1/3 $ at    the isotherm $T = 18$ MeV there exists the 2-nd order PT (dashed curve). }
 \label{Fig2} 
\end{figure}

\begin{figure}[ht]
\centering
\includegraphics[width=14cm, height=10cm]{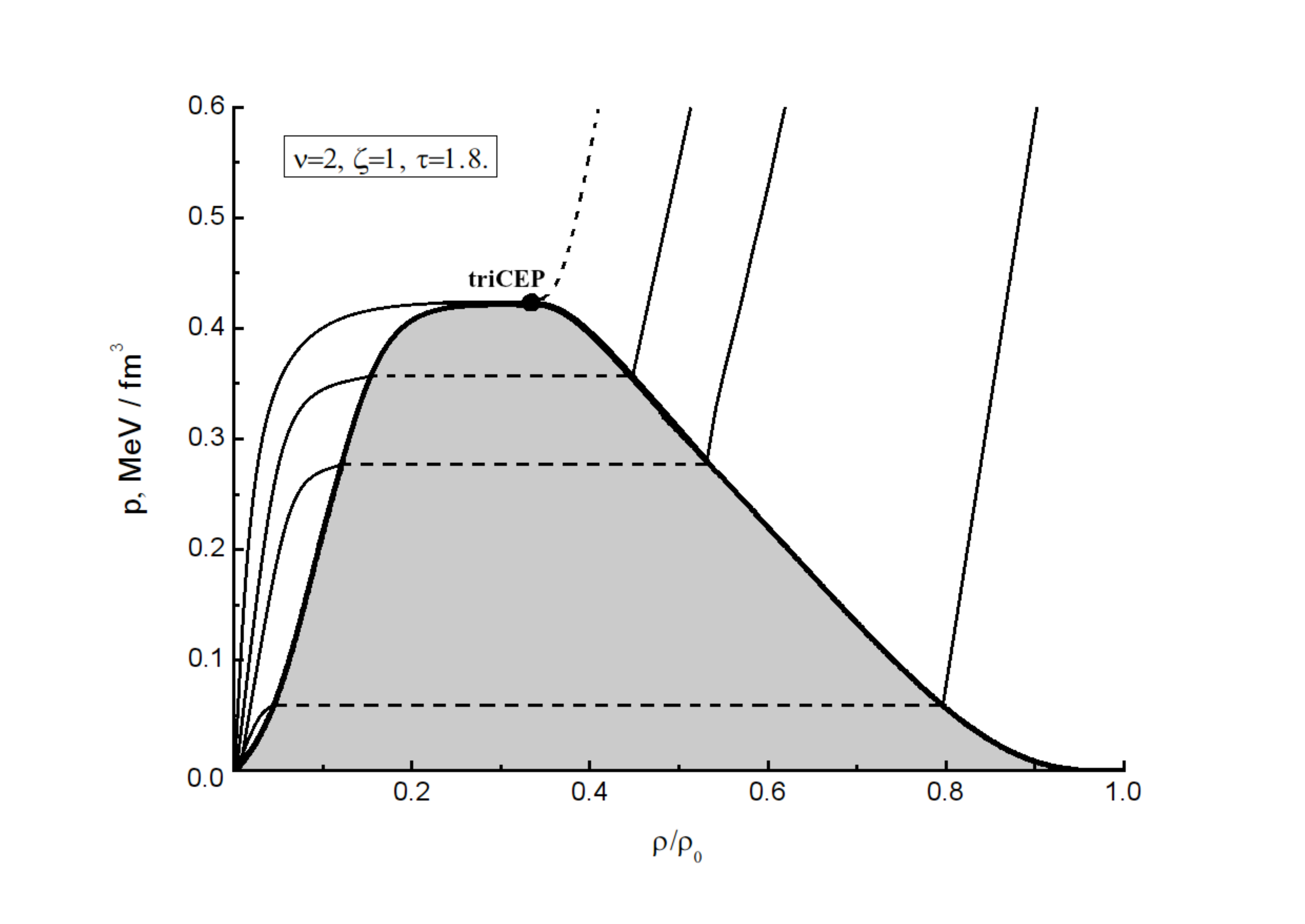}
\caption{The same as in Fig. \ref{Fig2}, but for the Fisher parameter  $\tau =1.8$.
 }
 \label{Fig2b} 
\end{figure}

A  careful analysis of  Eqs. (\ref{EqXII}) and  (\ref{EqXVII}) shows that for any integer $\nu > 1$ at $T_{cep}$ there is 
a curve of  the 2-nd order PT that always  begins at  the line of the 1-st order PT (see Figs. \ref{Fig1}-\ref{Fig2b}).  Therefore, all the critical endpoints of this model are, in fact, the tricritical  points.  This feature of the present model is similar to the simplified SMM solution for $ 1< \tau \le 2$ \cite{Bugaev:00, Reuter:01}, but its location  is  entirely different. 
 In the simplified SMM the 2-nd order PT  always exists  at  the particle number  density $\rho =\rho_0$ and for temperatures $T \ge T_{cep} = 18$ MeV, whereas in the present model such a PT occurs at the isotherm  $T = T_{cep} = 18$ MeV (see Figs. \ref{Fig1} and \ref{Fig1b})  along which the particle number density changes with  the pressure, as one can see from 
Figs. \ref{Fig2} and \ref{Fig2b}.   As one can see from Fig. \ref{Fig1b} a  variation  of  the parameter  $\tau$  from  $1.8$  to $2.2$ leads to  a negligible change of the 1-st order PT curve in  the $T-\mu$ plane  and to a  small variation of the nil curve of  surface tension  coefficient. 
As one can seen from Fig. \ref{Fig1b},  in  the limiting case $\tau \rightarrow \infty$ the nil curve of  surface tension  coefficient becomes an isotherm  $T=T_{cep}$.

Note that the shape of  phase diagrams of the present model in the $\rho-p$ plane shown in  Figs. \ref{Fig2} and \ref{Fig2b} look very similar to that ones for real liquids \cite{Stanley:71}, although   the critical isotherms  at the vicinity of the  tricritical  endpoint  shown in Figs. \ref{Fig2} and \ref{Fig2b}  are  rather flat compared to the mean-field calculations of Ref.  \cite{Satarov:2009zx}.  This  is due to the different values of the critical  index $\delta$ which in the mean-field models is $\delta_{mf} = 3$
\cite{Stanley:71} whereas in the present model  this index can be roughly  estimated as $\delta \simeq 6$ for $\tau =1.9$ and $\delta \simeq 5$ for $\tau =1.8$.  A more accurate
estimate for the index  $\delta$ can be provided by  the analytical  calculations of the critical exponents,  which is out of the scope of the present paper.

To constrain the choice of model parameters below we apply the L. van Hove axioms of  statistical mechanics \cite{Hove}, which shortly can be formulated as follows. Suppose that we are able to evaluate the exact  $N$ particle partition function $Q_N (V,T)$ of the given system.
\begin{enumerate}
\item Then in the thermodynamic limit, i.e. when  $N \rightarrow \infty$ and  $V \rightarrow \infty$ while the density  $\rho= N/V$ stays constant, the thermodynamic pressure  defined as 
\begin{eqnarray}\label{AxiomI}
p (\rho, T)= T \left( \frac{\partial \ln  Q_N}{\partial  V}  \right)_{N, T} \,, 
\end{eqnarray}
is a  {\it strictly non-negative  quantity.}
\item  The slope  $\left( \frac{\partial p}{\partial \rho} \right)_T$ of  any isotherm in $(\rho, p)$ plane  defined by  (\ref{AxiomI}) is {\it never negative.}  The limiting case corresponds to an existence of  ``flat'' region (regions) in which  $\left( \frac{\partial p}{\partial \rho} \right)_T = 0$ and, hence, the system becomes infinitely compressible.  The existence of such regions  in the  $(\rho, p)$ plane corresponds to the coexistence of two or more phases in the given system. 
\item The presence of the absolutely flat part of an isotherm with $\left( \frac{\partial p}{\partial \rho} \right)_T \equiv 0$ with the mathematical singularities at its ends, is a consequence of the limit $N \rightarrow \infty$.
If $N$ were finite, but large, then the pressure  (\ref{AxiomI})  {\it would be free from mathematical singularities}.  The usual sharp corners of an isotherm  would be round off and, at the same time, the usual flat part of the isotherm would not be really flat, but  instead it would have a small  positive slope $\left( \frac{\partial p}{\partial \rho} \right)_T$. 
\end{enumerate}
These axioms play an important role for the approximate partition functions too.  For example, 
 just the second axiom above  requires to use the Maxwell construction \cite{Stanley:71,Hove} in order to get rid of the isotherms of the Van der Waals type for which $\left( \frac{\partial p}{\partial \rho} \right)_T < 0$ in the phase transition region.  As we demonstrate below the present model obeys these axioms for some choice of the model parameters.

For a numerical evaluation of Eqs.  (\ref{EqXII}) and  (\ref{EqXVII}) we used the following parameterization of the eigen surface free energy  coefficient $ \sigma_{0} (T) $ 
\begin{eqnarray}
\label{EqXXIX}
\sigma_{0} (T) =  \sigma_{01}-\sigma_{02}\left(\frac{T}{T_{cep}}\right)^{\zeta}  \, ,
\end{eqnarray}
where the eigen surface tension at $T=0$ is taken from the standard SMM parameterization  $\sigma_{01}=18$ MeV and   $\sigma_{02}>0$ and  $\zeta$ are some  constants.   
  We found that for the parameter $\zeta \ge 1$ in the vicinity of the tricritical point  the total surface tension coefficient always vanishes as the first power of the difference $T-T_{cep}$, i.e. the temperature dependence of the surface tension of the 
present model is similar to the FDM  \cite{Fisher:67} and to the exactly solvable model of surface deformations  \cite{Bugaev:04a, Bugaev:07b}. 
Therefore, the value $\zeta = 1$ was used in actual calculations, while  the constant $\sigma_{02}$
was determined from Eq. (\ref{EqXVII}) to provide the critical endpoint existence at $\xi_2 =0$. 
For the parameter $\alpha =1.5$
used in our  calculations 
we find  $\sigma_{02} \simeq 24.76$ MeV for any $\nu > 1$.  Thus, at the critical endpoint Eq. (\ref{EqXVII}) can be explicitly written as 
\begin{eqnarray}
\label{EqXXX}
3\, V_{1} {I}_{\tau-\frac{1}{3}}\left(0, (1-\alpha) \sigma_0 (T_{cep})   \right) = - \sigma_{0} (T_{cep}) \,.
\end{eqnarray}
On the one hand, this equation can be considered as a condition to determine $\sigma_{02}$ for  a given value of $T_{cep}$, but, on the other hand, for a fixed value of  the constant $\sigma_{02}$ it can be regarded as an  equation for  unknown value of 
$T_{cep}$.  Our numerical  analysis shows that  for $\nu=2$ and  $\nu=4$ there exists a single solution for $T_{cep}$. 

For $\nu =3$ Eq. (\ref{EqXXVII}) has three branches of  solutions at $T_{cep} = 18$ MeV which correspond to  three tricritical points  at this temperature, but only one of them is a physical one. 
The solution  with the   lowest  value of chemical potential $\mu_{cep}$  which  corresponds to the negative particle density  is  similar to the corresponding PT curves  for  $\nu =2$.  The two remaining solution have, respectively, either  negative   values   of the chemical  potential $\mu_c (T) \le - W_0 $ along the coexistence 
or  positive values of  $\mu_c (T) > 0$ for $T \le T_{cep}$. 
 In order to understand which solution   
is  the physical  one,  we  apply   the second L. van Hove  axiom given above.   The latter requires  that   the inverse compressibility modulus   $1/K_{T}$ of liquid phase alone  should   not be negative. From this  requirement one gets 
\begin{eqnarray}
\label{EqXXXI}
K_{T}\equiv\frac{\nu(\nu-1)}{V_{1}\rho^{2}_{L} \tilde \mu^{2}}\cdot a_{\nu} \tilde\mu^{\nu} \ge 0
\quad 
\Longrightarrow 
\quad 
 a_{\nu}\tilde \mu^{\nu} \ge 0 .
\end{eqnarray}
Since  for $\nu=3$  the condition $\rho_{cep} \simeq \rho_0/3$ can be  obeyed  only for    negative value of the  parameter  $a_{3}$, i.e.  $a_{3}<0$, then    according to (\ref{EqXXXI})
the positive compressibility of this liquid corresponds to the inequality $\tilde \mu \le  0$, i.e. $\mu \le  - W_0$.  
Therefore, any  additional   tricritical point  that  corresponds to  positive value of chemical potential $\mu$
is   unphysical. 
 This indicates that 
for $\nu =3$ the liquid phase pressure (\ref{EqXXIV}) is not very realistic, since it is not able to describe the observed 
experimental states at high particle number densities.  Actually, the same arguments are applicable to the odd 
powers $\nu \ge 3$ and, hence,  the corresponding EoS  are also unphysical.

This is also  the case for the even values of power  $\nu$ in  (\ref{EqXXIV}), except for  
the case $\nu =2$. This follows from the 
 analysis of  the nuclear incompressibility modulus, $K_0 \equiv 9  \left(  \frac{\partial p_L}{\partial \rho_L} \right)_{T} $,  at normal nuclear  density and vanishing temperature.   Indeed, from (\ref{EqXXIV}) one can find  explicitly 
\begin{eqnarray}
\label{EqXXXII}
K_0 = \frac{9}{\nu(\nu-1) \, a_{\nu}\,  \tilde\mu^{\nu-2}}  \,. 
\end{eqnarray}
This expression shows that for the adopted parameterization of  liquid phase EoS the coefficient  $K_0$ is finite only for 
$\nu =2$ and $K_0 (\nu=2)  = \frac{9}{2 \, a_{2}} \simeq 357$ MeV, while for  any power $\nu > 2$ the incompressibility modulus  (\ref{EqXXXII})  diverges, because  $\tilde\mu = 0$ at $T=0$ (for more details see Appendix A).
Experimental values of  the nuclear incompressibility modulus  are quoted as  $K_0^{exp} \simeq 230 \pm 30$ MeV \cite{Kfactor:1, Kfactor:2,Khan:2009}, but theoretical  models with larger  value up to    $350-380$ MeV
are also  known.  For instance, the Skyrme force model SIII, which is able to  well describe  the empirical  properties  of many nuclei \cite{Kfactor:3}, has  the value of the  incompressibility modulus  $K_0 \simeq 355 $ MeV. 

The obtained value of  $K_0 (\nu=2)$ can, of course,  be reduced to $K_0 (\nu=2) \simeq 250$ MeV by choosing $\mu_0 \simeq - 6$ MeV in  the parameterization  (\ref{EqAXXXII}), but this immediately leads to a strong  decrease of the particle number  density of the liquid at the phase equilibrium curve at $T=0$ to $\rho_L \simeq 0.7 \rho_0$ which is much more unrealistic than the large value of the  nuclear incompressibility modulus.

From the discussion above it is clear that only  the liquid phase pressure (\ref{EqXXIV}) with $\nu = 2$  fulfills the second L. van Hove  axiom. A close inspection shows that the incompressibility modulus of the gaseous phase is always positive (see also \cite{Bugaev:00}). Since the present model is an exactly solvable, it automatically leads to  the Gibbs criterion of phase equilibrium \cite{Bugaev:00,Bugaev:00b,Reuter:01,Bugaev:07,Bugaev:07a}.  Moreover, in the mixed phase  the pressure  derivative  with respect to  $\rho$ at constant $T$ is exactly zero, i.e. $\left(\frac{\partial p}{\partial \rho}\right)_{T}= 0$, i.e. the proposed model EoS has no regions of negative $\left(\frac{\partial p}{\partial \rho}\right)_{T}$ values
  (see isotherms in  Figs. \ref{Fig2} and \ref{Fig2b}). 
Therefore, the present model EoS obeys the second  axiom of  statistical mechanics \cite{Hove} everywhere.
 Although the liquid phase pressure (\ref{EqXXIV}) can be negative, but  the gaseous phase  pressure  is always positive and, hence, in this region the phase equilibrium cannot be achieved and in this case the gaseous phase dominates. Therefore, this EoS obeys the first axiom of  statistical mechanics \cite{Hove} too.  
We would like to stress that the present model  has the same physical mechanism of the PT generation as the simplified  SMM  \cite{Bugaev:00, Bugaev:00b, Bugaev:07} and, hence, it can be solved analytically for finite volumes 
using the methods suggested in \cite{Bugaev:05}. For finite volumes the number of sorts of fragments $N$ in Eqs.  (\ref{EqX}) and  (\ref{EqXI}) is restricted from above $N \le V/V_1$ and because of that the essential  singularity of the isobaric partition 
 disappears \cite{Bugaev:05} and  a  PT is washed out \cite{Bugaev:05,Bugaev:07}.
Moreover, for finite, but large number of particles $N$  inside the finite volume analog of mixed phase  the derivative $ \left(\frac{\partial p}{\partial \rho}\right)_{T}$ of  thermodynamic pressure (\ref{AxiomI}) vanishes  
as $c^2/N$. 
Therefore, the present model EoS obeys the third L. van Hove axiom \cite{Hove}  as well.


\section{Manifestation of  negative surface tension values}

\begin{figure}[ht]
\centering
\includegraphics[width=16cm, height=10cm]{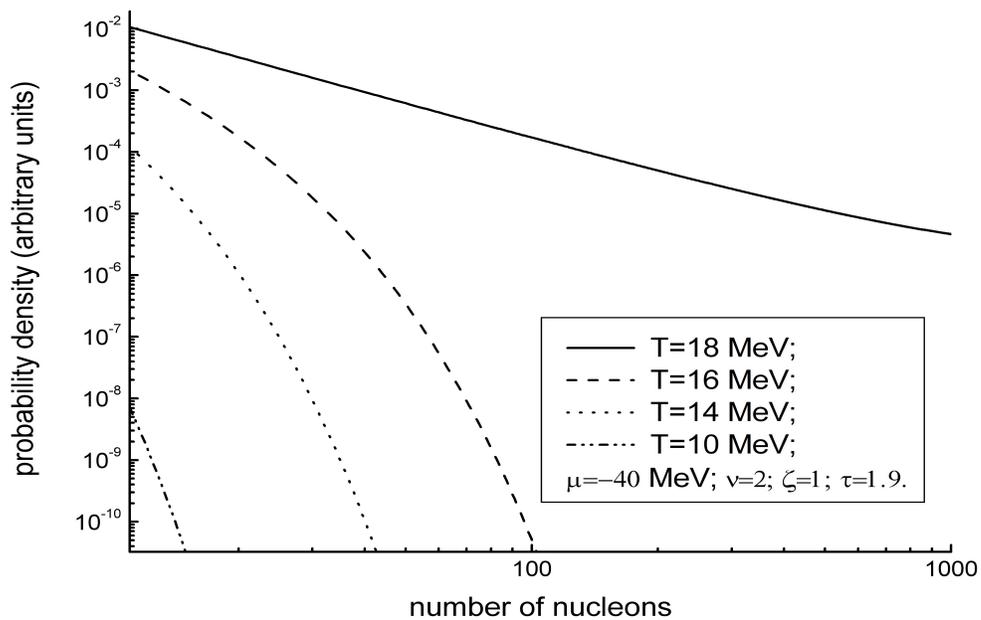}
\caption{Fragment size distribution in the gaseous phase  is shown for different temperatures and fixed baryonic chemical potential and $\nu=2$. For all temperatures below $T_{cep} = 18$ MeV the surface tension coefficient is positive, while it vanishes for $T = T_{cep}$.
As temperature increases the distribution changes from the exponential one to a power law which is a straight line in a double logarithmic scale.}
\label{FigD_1}
\end{figure}

In this section we  study the fragment size distributions in different parts of the phase diagram in order to elucidate 
the role of the negative surface tension coefficient.  But first we consider the gas of nuclear fragments with nonnegative surface 
tension coefficient, i.e. for  the  dimensionless surface term $\xi_2 \ge 0$.  In this case the fragment size distribution is proportional to 
\begin{eqnarray}
\label{EqNXXXIII}
\omega_{gas} (k) = \exp \left[ - \xi_1 \, k - |\xi_2|k^ \varkappa - \tau \ln k \, \right] \,, 
\end{eqnarray}
where  $k$ is the number of nucleons in a fragment.
It is monotonically decreasing function of  the  fragment size $k$, since  $\xi_1 > 0 $ is  valid everywhere 
except for the PT curve, where   $\xi_1 = 0 $.  As one can see  from Fig. \ref{FigD_1},  for  all temperatures below the critical one the  low $k$  behavior   is governed by the Fisher exponent $\tau$ and by the surface term which is proportional to $\xi_2$, whereas  at large values of $k$ the fragment size attenuation is solely determined  by the bulk term  which is proportional to  $\xi_1$. These  fragment size  distributions are qualitatively similar  to that ones found recently in  \cite{SMM:12} for  the nuclear  matter  EoS neutralized by the electrons in the stellar environments.
A true power law  of the  fragment size distribution corresponds to the tricritical endpoint, for $\xi_1= 0$  and $\xi_2 = 0$. 
In Fig.  \ref{FigD_1} one can also see the  effective power law  of the  fragment size distributions  for temperatures which are very close  to $T_{cep}$.  A similar behavior can be seen at lower temperatures, if  $\xi_1 \ll 1$ and $\xi_2 \ll 1$.

\begin{figure}[ht]
\centering
\includegraphics[width=16cm, height=10cm]{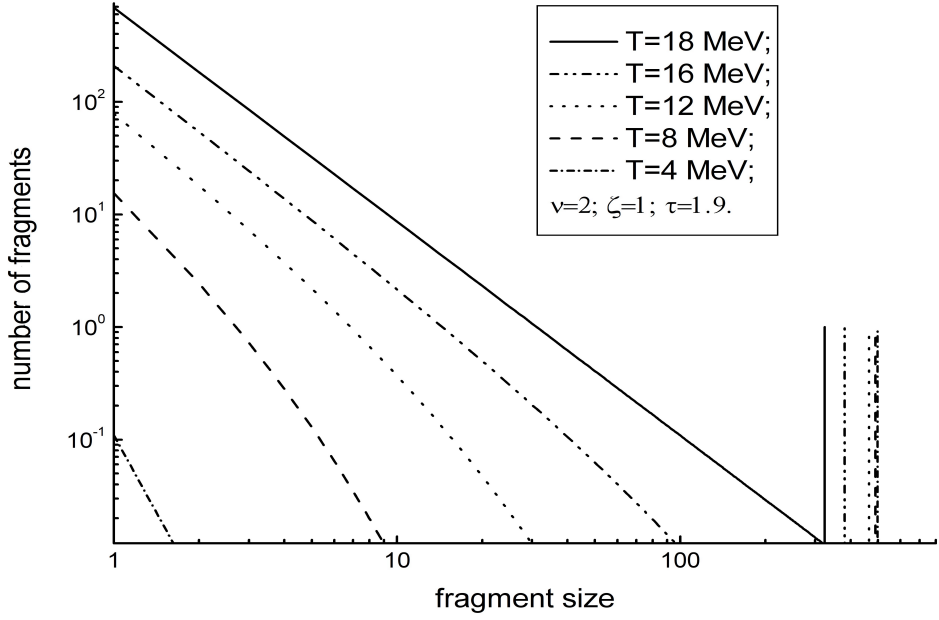}
\caption{Fragment size distribution in the mixed phase  is shown for different temperatures and fixed particle number  density $\rho = \frac{\rho_0}{3}$ and $\nu=2$. For all temperatures below $T_{cep} = 18$ MeV the surface tension coefficient is positive, while it vanishes for $T = T_{cep}$. 
All these distributions were normalized for 500 nucleons in the system. The liquid cluster is represented by a Kronecker delta function. 
As temperature increases the distribution function  of the gas  changes  from the exponential one to a power law which is a straight line in a double logarithmic scale.}
\label{FigD_2}
\end{figure}

In the mixed phase the bulk term vanishes due to the PT condition $\xi_1 = 0$ and, hence, the fragment distribution 
of  the gas phase acquires the form
\begin{eqnarray}
\label{EqNXXXIV}
\omega_{gas}^M (k) = \exp \left[ - |\xi_2|k^ \varkappa - \tau \ln k \, \right] \,,
\end{eqnarray}
which is governed by the Fisher  term (at small $k$ only)  and by  the surface term  (see Fig. \ref{FigD_2}).  In addition to the gaseous phase distribution (\ref{EqNXXXIV})
the mixed phase contains an infinite fragment representing the liquid phase. Its distribution function corresponds to the 
Kronecker  $\delta$-function. In order to make a comparison with the results reported in \cite{SMM:12} we have normalized the 
fragment size distributions in the mixed phase to the total number of    500  nucleons. In  Fig. \ref{FigD_2} we present results for 
 the fixed  particle number  density $\rho = \frac{\rho _0}{3}$. As one can see,  at low temperatures  the gas of fragments is practically absent and, hence, almost  whole matter belongs to the liquid fragment with the size of 
almost 500 nucleons.  For higher temperatures the size of liquid fragment decreases and  the gaseous phase appears. 
At the tricritical point about 65 \% of  the  matter belongs to the liquid fragment, while the rest  belongs to the gas, which demonstrates a clear power law.   Such a behavior of the  fragment size 
distributions in the mixed phase is qualitatively similar to the one  found in  \cite{SMM:12} inside the mixed phase, 
see Figs. 4 and 6 in \cite{SMM:12}. The main difference with the results of  \cite{SMM:12} is that the present model 
employes the grand canonical ensemble in thermodynamic limit, whereas \cite{SMM:12} is dealing with the canonical ensemble
for a finite system. In the latter  case the size of largest fragment, which represents the liquid phase,  fluctuates from  event to event and, hence, instead of the Kronecker  $\delta$-function distribution one gets a Gaussian one with a finite width.

A principally new type of  the fragment size attenuation corresponds to the region with  negative values of the 
total surface tension coefficient, i.e. for $\xi_2 < 0$.  In this case  the unnormalized distribution of nuclear fragments 
has the form 
\begin{eqnarray}
\label{EqXXXIII}
\omega (k) = \exp \left[ - \xi_1 \, k + |\xi_2|k^ \varkappa - \tau \ln k \, \right] \,,
\end{eqnarray}
i.e. it
has the local minimum at some value $k_{min}$ and the local maximum at  $k_{max} > k_{min}$. This can be shown by inspecting the logarithmic derivative of $\omega(k)$ with respect to $k$. The extremum condition for such a derivative is
given by the equation 
\begin{eqnarray}
\label{EqXXXIV}
\left. \frac{\partial \ln \omega (k)}{\partial \, k}\right|_{k=k_E} = - \xi_1  + \varkappa\, \frac{ |\xi_2|}{k^{1-\varkappa}_E}\,   - \frac{\tau}{k_E}  = 0 
\quad \Rightarrow \quad k_E =  \left[ \frac{\varkappa\,  |\xi_2| }{\xi_1 + \frac{\tau}{k_E }}\right]^\frac{1}{1- \varkappa}
\,,
\end{eqnarray}
where the extremum is reached for $k = k_E$.  Let us show now that the expression for $k_E$ in (\ref{EqXXXIV}) has two positive solutions. In first case we assume that $\xi_1 \ll  \frac{\tau}{k_E }$, which may occur  for  small values of  $k_E$. Then neglecting the term  $\xi_1$ in the above expression  for  $k_E$ one finds 
\begin{eqnarray}
\label{EqXXXV}
 k_{min}  \simeq   \left[ \frac{\tau   }{\varkappa\,  |\xi_2|}\right]^\frac{1}{\varkappa}
\,. 
\end{eqnarray}
The analysis of the second derivative of $\ln \omega(k)$ with respect to $k$ 
\begin{eqnarray}
\label{EqXXXVI}
\left. \frac{\partial^2 \ln \omega (k)}{\partial \, k^2}\right|_{k=k_{min}} =  - \varkappa (1 -  \varkappa)\, \frac{ |\xi_2|}{k^{2-\varkappa}_{min}}\,   + \frac{\tau}{k^2_{min}}  =   \frac{ \varkappa \,\tau}{k^2_{min}} > 0 
\,,
\end{eqnarray}
shows that this derivative is always positive, i.e. there is a local minimum, for $\varkappa > 0$.  Note that Eq. 
(\ref{EqXXXV})  allows one to estimate  the dimensionless  surface term   $\xi_2 \simeq  -  \frac{\tau}{\varkappa\, k_{min}^\varkappa}$, if the  position of  the local minim  is known.

\begin{figure}[ht]
\centering
\includegraphics[width=16cm, height=10cm]{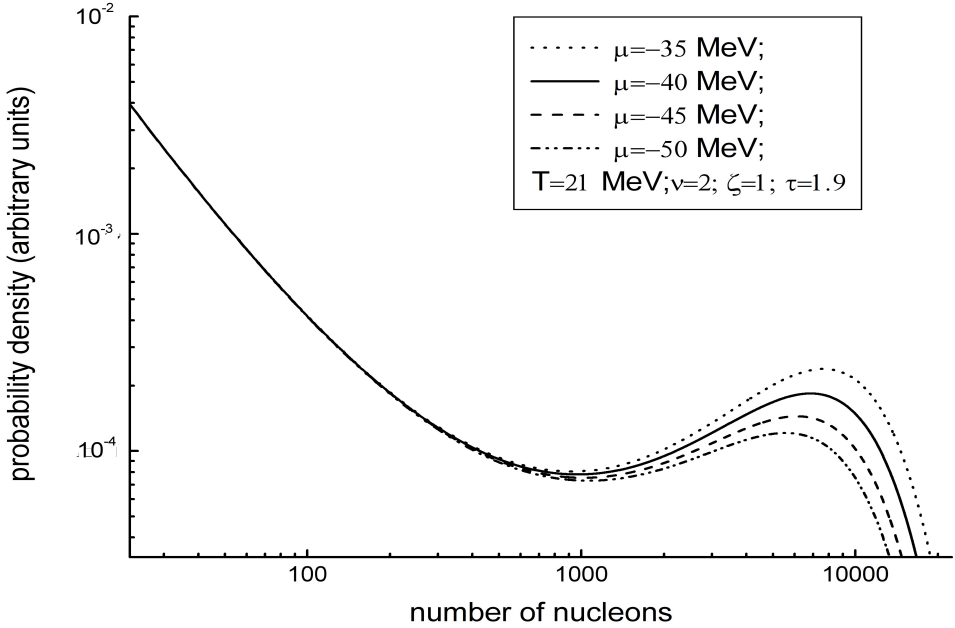}
\caption{Fragment size distribution in the phase with the negative value of the surface tension  coefficient  of the model with $\nu=2$  is shown for a fixed  temperature 
and different baryonic chemical potentials. 
As baryonic chemical potential  increases the maximum of  the distribution function  grows  higher and wider,  and it shifts 
towards the  larger 
number  of nucleons in a fragment.}
\label{FigD_3}
\end{figure}

\begin{figure}[ht]
\centering
\includegraphics[width=16cm, height=10cm]{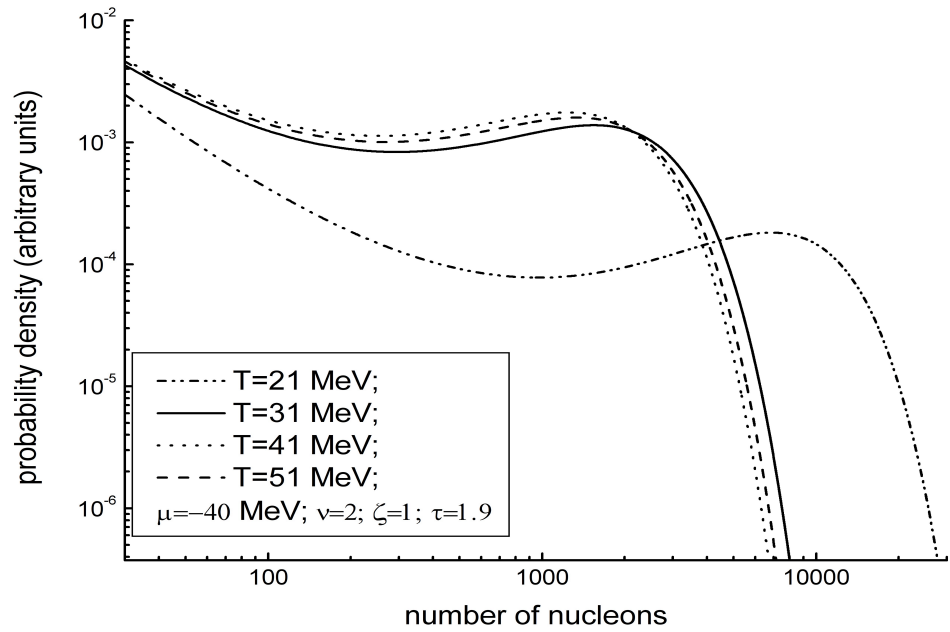}
\caption{Fragment size distribution in the phase with the negative value of the surface tension  coefficient  of the model with $\nu=2$  is shown for a fixed baryonic chemical potential   
and different  temperatures. 
As temperature   increases the minimum and  maximum of  the distribution function  grow   wider and shallower and they shift towards the  smaller 
number  of nucleons in a fragment.}
\label{FigD_4}
\end{figure}

In the opposite case, if  $\xi_1 \gg  \frac{\tau}{k_E }$, which  occurs  for  large values of  $k_E$, the solution for $k_E$
takes the form 
\begin{eqnarray}
\label{EqXXXVII}
 k_{max}   \simeq   \left[ \frac{\varkappa\,  |\xi_2|  }{ \xi_1}\right]^\frac{1}{1-\varkappa}
\,.
\end{eqnarray}
The second derivative of $\ln \omega(k)$ with respect to $k$  can be written as 
\begin{eqnarray}
\label{EqXXXVIII}
\left. \frac{\partial^2 \ln \omega (k)}{\partial \, k^2}\right|_{k=k_{max}} =  - \varkappa (1 -  \varkappa)\, \frac{ |\xi_2|}{k^{2-\varkappa}_{max}}\,   + \frac{\tau}{k^2_{max}}  =  -  \frac{1}{k_{max}}  \left[ \xi_1(1 -  \varkappa) -   \frac{\tau}{k_{max}}\right]
\,.
\end{eqnarray}
This  derivative  is  negative for   $\xi_1(1 -  \varkappa) > \frac{\tau}{k_{max}}$.  This  inequality cannot be fulfilled only for $(1 -  \varkappa) \ll 1$, whereas for the typical 
SMM value $\varkappa \simeq \frac{2}{3}$ it  is fulfilled  due to 
adopted  assumption  $\xi_1 \gg  \frac{\tau}{k_{max} }$. Thus, at $k \simeq k_{max}$ the fragment distribution (\ref{EqXXXIII}) has a local maximum. 
The size distributions  with the saddle-like shape which  have   both  a local minimum and  a local maximum  are clearly seen in   Figs. (\ref{FigD_3}) and (\ref{FigD_4}). 
Our analytical estimates  are well supported by the numerics. 

Combining the  expressions (\ref{EqXXXV}) and (\ref{EqXXXVII}), one can  get an approximate relation between the mass  numbers  of fragments that correspond to these two extrema 
\begin{eqnarray}
\label{EqXXXIX}
 k_{min}^\varkappa \,   k_{max}^{1 -  \varkappa}    \simeq  \frac{\tau   }{\xi_1} \,,
\end{eqnarray}
which allows one to estimate the dimensionless bulk term $\xi_1$ for a given fragment mass distribution which has two  extrema. 

Another distinctive feature of the fragment size distributions with the negative surface tension coefficient is a presence of 
a quasi-power law for a wide range   of  fragment sizes below $k_{min}$.  It appears due to the fact that 
$ k_{min}$ and  $k_{max}$  are large  because of  small values   
of bulk $\xi_1$ and  surface  $\xi_2$ terms.
 If the fragment size exceeds $k_{max}$, then the  bulk term  in  (\ref{EqXXXIII}) dominates and the distribution
$\omega(k)$ becomes  exponential,  which in  a double logarithmic scale looks like    a fast fall off.

Note that a quasi-power law persists to exist even, if both a minim and a maximum move to each other   and  become very shallow. 
The condition of their disappearance is just $k_{min} = k_{max}$, which with the help of  (\ref{EqXXXIX}) can be represented  as 
\begin{eqnarray}
\label{EqXXXX}
\frac{\tau   }{\xi_1} \simeq  \left[  \frac{\tau   }{ \varkappa \, |\xi_2|}  \right]
 ^\frac{1}{\varkappa }    \,.
\end{eqnarray}
In this case a quasi-power law at small values of $k$ changes to a slightly convex shape which at larger   values of $k$  acquires a 
strong concave shape and  then it turns into an exponential all off. From (\ref{EqXXXX}) it is easy to find that the reason for such a behavior is just a cancellation of the bulk  and surface contributions in the vicinity of 
$k = k_{min} = k_{max}$.

It is necessary to mention  that  the saddle-like mass distributions were reported in  Ref. \cite{Campi:03} where  the authors numerically studied 
the multifragmentation scenario for the ensemble of  classical particles interacting via the Lennard-Jones potential.  
Thus, the full  curve in Fig. 2
of   \cite{Campi:03} is very similar to the distributions with the negative surface tension coefficient 
discussed above, 
 although  its extrema are not  so well pronounced as the ones shown in  Figs.  \ref{FigD_3} and \ref{FigD_4}. 
It is, of  corse, difficult to directly compare our results for the fragment size distributions  with the ones found in  \cite{Campi:03} and to make some definite conclusions, since the authors of \cite{Campi:03}  either studied small systems (189 particles) in which the finite size effects are strong (see Fig. 2 in  \cite{Campi:03}) or they have low statistics for the power law
shown in Fig. 8 of  \cite{Campi:03}. Nevertheless,  the authors of  \cite{Campi:03} claimed  to  find  a curve of 
power law  size attenuations with 
nearly constant energy per particle. 
Note that
such a behavior is typical for  the present model along the curve of the 2-nd order PT.  In \cite{Campi:03} it was also found that  in the course of expansion from the initially   dense state to a dilute  state the fragments are highly nonspherical and they have fractal surfaces.  The above analysis of the gaseous  fragment size distributions shows us that the fractals which  appear at  the tricritical point or at the states of the 2-nd order PT  correspond to  a power law dependence  on the size of fragments, whereas the fractals associated 
with the negative values of the surface tension coefficient manifest themselves in the nonmonotonic fragment size distributions  
of  the saddle-like shape. 
Therefore, it is quite possible that the negative surface tension is responsible for the peculiar and nonmonotonic  fragment size 
distributions reported in \cite{Campi:03}, but for a more definite conclusion a detailed comparison between the  two models  is needed. 

\section{Conclusions}\label{secConclusions} 

In the present work we propose a new SMM formulation based on the consistent 
treatment  of the second virial coefficients for the ensemble of  nuclear fragments.  
Such a  virial expansion  allows us to explicitly  account  for  the many-body  effects.
Our analysis shows that interaction between the nuclear fragments induces  an  additional contribution into the surface tension   free energy.
 It is shown that by a proper choice of the temperature dependence of the full surface tension the standard SMM  which accounts 
 only  for   the proper volumes of fragments   is able to correctly  reproduce  the low density  virial expansion up to the second order.   This resolves an old puzzle of why the SMM is so good at low 
 densities, although it employes the hard core repulsion approximation  which is suited for high densities only.
 
 The present model, however,  leads to an additional equation for the induced surface tension coefficient, which at the moment accounts only  for the repulsion between the  nuclear fragments. In order to simplify the presentation of our idea  and  to make easier  an analysis of the model phase diagram,  here   we assumed  that the effects of   attraction between the nuclear fragments are  implicitly accounted  in the temperature dependent  surface tension coefficient of   the fragments.  However, 
the treatment of the model with an  attraction between the fragments will be our  next step.

Another important result of the present work  is extension of the previous model for the case of a compressible nuclear liquid. This is achieved by introducing    an additional  $\mu$-dependent term in the expression for liquid phase pressure. Note that the obtained  model  EoS for $\nu =2$ obeys the basic axioms of the statistical mechanics 
formulated by L. van Hove.
It does not lead to an appearance of the non-monotonic isotherms in the mixed phase region which are typical for the mean-field models. A direct consequence  of the finite  liquid compressibility is that  the present model allows us  to generate the tricritical endpoint at $\rho_{cep} = \rho_0/3$,  which is a typical  value of critical density  for  the liquid-gas PTs in the  ordinary liquids.  This  novel feature makes the present model more realistic than the standard SMM.

 The range of the Fisher parameter in this model is $\tau \ge 1$, whereas in the simplified SMM all  values of  the parameter  $\tau$ are allowed.  The other important difference with the simplified SMM is that in the present model there exist the tricritical points only and this is possible  for  $1 \le \tau \le 2$.  Thus,  our analysis  showed that  in the present  model  each  1-st order PT curve is ended at the curve of the 2-nd (or higher) order PT. 
  It is found  that the 2-nd order PT in the tricritical point exists for  $\frac{3}{2} \le \tau \le 2$, while  the $n$-th  ($n =3, 4, 5, ...$) order PT  in a  tricritical point    exist for $\frac{n+1}{n} \le \tau < \frac{n}{n-1}$.  
    
 For  $\tau > 2$ we found that, in contrast to the simplified SMM, the 1-st order PT exists at $T \le T_{cep}$ and, hence,  in  this case instead of the critical endpoint the model exhibit a critical line. Therefore, the physical range of the Fisher parameter $\tau$ should be constrained by   $\frac{3}{2} \le \tau \le 2$.

Our  analysis of the fragment size distributions  in  the region of   negative  surface tension coefficient showed that these distributions have an unusual saddle-like shape. We established a simple relation between the local minimum position  in the fragment  size $k_{min}$ and the dimensionless  surface term $\xi_2 \simeq  -  \frac{\tau}{\varkappa\, k_{min}^\varkappa}$ which can be used to estimate the $\xi_2$ value  directly from the fragment size distribution. 
We would like to stress that  the distributions of  a  similar shape were observed  in molecular dynamical  studies  of multifragmentation for the particles interacting via the Lennard-Jones potential \cite{Campi:03}. 

Highly  nonspherical shapes of the  fragments  observed in these simulations, allow us to believe  that all these features of the dynamical  multifragmentation can be related to the negative surface tension coefficient in supercritical region. 

In the developed model  the contribution of  surface tension induced by the repulsive  interaction between the nuclear fragments is evaluated  in a simplified form. Nevertheless,  in  combination with a  finite   incompressibility of  liquid phase  it gave us  rather rich phase structure of the nuclear matter phase diagram. 
It is clear that a realistic short-range   attraction   between the nuclear  fragments should also be included  in order to correctly  locate  the region of the 2-nd order phase transition. 

\vspace*{4mm}

{\bf Acknowledgments.} 
The authors appreciate the valuable comments of  L. M. Satarov. 
V.V.S., A.I.I. and K.A.B.  acknowledge  a partial  support of the Program `On Perspective Fundamental Research in High Energy and Nuclear Physics' launched by the Section of Nuclear Physics  of National Academy of  Sciences of Ukraine. 
K.A.B.  and I.N.M.  acknowledge a partial support provided by the Helmholtz 
International Center for FAIR within the framework of the LOEWE 
program launched by the State of Hesse. The work of   I.N.M. was also supported in part by the grant NSH-215.2012.2 (Russia).


\section{Appendix A}

In this Appendix we motivate for the parameterization  of the liquid phase pressure (\ref{EqXXIV}).
For this purpose, first, we consider a more general choice of the liquid phase pressure
\begin{eqnarray}
\label{EqAXXXII}
p_L=\frac{\tilde W + \mu - \mu_0 +a_{\nu}\left[ \mu  - \mu_0 \right]^{\nu}}{V_1}\, ,  \quad {\rm for} \quad \nu = 2, 3, 4 \,,
\end{eqnarray}
where $\tilde W(T)$ and $\mu_0$ are assumed to be   the functions of $T$, i.e. $\tilde W(T) $ and  $\mu_0 (T)$. For  the particle number  $\rho_{L}$ and entropy $s_{L}$ densities of  liquid  phase one finds  
\begin{eqnarray}
\label{EqAXXXIII}
\rho_{L} & \equiv &  \frac{\partial\, p_L}{\partial \,\,\mu} =\frac{1+a_{\nu}\nu \tilde \mu^{\nu-1}}{V_{1}} \,, \quad {\rm with} \quad  \tilde\mu \equiv \mu  - \mu_0\, , \\  
\label{EqAXXXIV}
s_{L} & \equiv &  \frac{\partial\, p_L}{\partial \,\, T} =\frac{1}{V_{1}} \left[ \frac{d \tilde W}{d T} -   \frac{d \tilde  \mu_0}{d T} \frac{\rho_L}{\rho_0}  \right] \,. 
\end{eqnarray}
In order to provide the  nonnegative values of  the liquid  entropy density    $s_{L} \ge 0$ at low and  high particle densities $\rho_L$, 
  it is necessary to require  that  $\mu_0 = const$. Then   from  (\ref{EqAXXXIV}) it is clearly seen that the liquid entropy density is positive for any large  densities $\rho_L \gg \rho_0$, if  $\mu_0 = const$ and if  $\frac{d \tilde W}{d T} \ge 0$.
Moreover, according to the third law of thermodynamics, the entropy density of the system must vanish  at $T=0$, i.e. 
$\frac{d \tilde W }{d T} |_{T=0} = 0$ which is fulfilled  automatically, 
 if  $ \tilde W(T) = W_{Fm} (T) + C$, where  $C$ is a constant to be found and 
$W_{Fm} (T) \equiv \frac{T^2}{\varepsilon_0}$ ($\varepsilon_0=16$ MeV) is the contribution of the excited states taken in the Fermi-gas approximation as in the original SMM  \cite{Bondorf:95}. 

In order to determine  the constant  $C$,  consider  the  Gibbs criterion of  phase equilibrium  $\xi_1(T=0, \mu)$ at $T=0$. Since at $T=0$ the gaseous pressure is zero, then  an  explicit form   of  such a criterions is as follows 
\begin{eqnarray}
\label{EqAXXXV}
\tilde W(0) + \tilde \mu  +a_{\nu} \tilde \mu^{\nu}  = 0\, .
\end{eqnarray}
From this algebraic equation for $\tilde \mu$ one  deduces  that for any real $\tilde W(0)$ value Eq. (\ref{EqAXXXV}) has exactly  $\nu$ algebraic roots. 
Analyzing  Eqs. (\ref{EqAXXXIII}) and (\ref{EqAXXXV}), one concludes that in order to have  the nuclear liquid 
of normal nuclear density $\rho_0 = V_1^{-1}$  at the phase equilibrium  point   $T=0$, it is necessary  that 
$\tilde \mu = 0$ is a solution of (\ref{EqAXXXV}).  
 Moreover, it is easy to see that this is the only physical solution.
 An existence of the solution  $\tilde \mu = 0$  of  (\ref{EqAXXXV}) is provided by the condition $\tilde W(0) = 0$.  
 This condition   along  with  the requirements for  $\frac{d \tilde W}{d T}$ found above, unambiguously leads to that $\tilde W(T) = W_{Fm}(T) = \frac{T^2}{\epsilon_0}$ and $C=0$,  and, hence,  $\mu_0 = - W_0 = - 16$ MeV. Such a choice not only obeys all the conditions discussed above, but also it automatically guarantees that in  the vicinity of  the normal nuclear state, i.e. at $T=0$ and $\rho_L = \rho_0$,   the liquid phase of the present model (and its phase diagram) coincides with that one of  the simplified SMM.



\newpage


\begin{thebibliography}{99}


\bibitem{Bondorf:95}
J. P. Bondorf et al., Phys. Rep. 257, 131 (1995) and references therein.

\bibitem{Gross:97}
D. H. E. Gross, Phys. Rep. {\bf 279}, 119 (1997).

\bibitem{Moretto:97}
L. G. Moretto {\it et al.},  Phys. Rep.  {\bf 287}, 249 (1997).

\bibitem{Randrup:04}
%
P. Chomaz, M. Colonna and J. Randrup, Phys. Rep.  {\bf 389}, 263 (2004). 

\bibitem{Bugaev:07}
%
K. A. Bugaev, 
Phys. Part. Nucl. {\bf 38},  (2007), 447.

\bibitem{Igor:2010a} 
%
A.~S.~Botvina and  I.~N.~Mishustin, 
Nucl.   Phys.\ A {\bf 843}, 98 (2010).

\bibitem{Igor:2013b}
%
N.~Buyukcizmeci {\it et. al.,}  Nucl.   Phys.\ A {\bf  907}, 13  (2013).

\bibitem{Igor:AstroC}
%
N.~Buyukcizmeci,  A.~S.~Botvina and  I.~N.~Mishustin,  arXive:1304.6741 [nucl-th].


  
 \bibitem{Mishustin:2006ka} 
  I.~N.~Mishustin,
  Eur.\ Phys.\ J.\ A {\bf 30}, 311 (2006).
  

\bibitem{Mishustin:1998eq} 
  I.~N.~Mishustin,
  Phys.\ Rev.\ Lett.\  {\bf 82}, 4779 (1999).

\bibitem{Mishustin:2008d}
%
G. Torrieri, B. Tomasik and    I. N. Mishustin,
 Phys.\ Rev.\ C  {\bf  77},  034903    (2008). 
 
 \bibitem{simpleSMM:1}
%
S. Das Gupta and A.Z. Mekjian, Phys. Rev. {\bf C 57}, 1361 (1998).

\bibitem{simpleSMM:1b}
%
S. Das Gupta, A. Majumder, S. Pratt, and A. Mekjian,  
nucl-th/9903007.

\bibitem{Bugaev:00}
K. A. Bugaev,
M. I. Gorenstein, I. N. Mishustin and W. Greiner,
Phys. Rev. {\bf C62},  044320 (2000);
arXiv:nucl-th/0007062 (2000).

\bibitem{Bugaev:00b}
K. A. Bugaev, M. I. Gorenstein, I. N. Mishustin and W. Greiner,
Phys. Lett. {\bf B 498},  144 (2001);
arXiv:nucl-th/0103075 (2001).


\bibitem{ISIS}
%
L. Beaulieu  {\it et al.,}  Phys. Lett. {\bf B 463}, 159 (1999).


\bibitem{EOS:00}
J. B. Elliott {\it et al.,} (The EOS Collaboration),
Phys. Rev. {\bf C 62},  064603 (2000).

\bibitem{Fisher:67}
M. E. Fisher, Physics {\bf 3},  255 (1967).

\bibitem{Elliott:06}
%
for a review on Fisher scaling  see  
J.~B.~Elliott, K.~A.~Bugaev, L.~G.~Moretto and L.~Phair,
arXiv:nucl-ex/0608022 (2006) 36 p. and references therein.

\bibitem{Reuter:01}
P. T. Reuter and K. A. Bugaev,
Phys.\ Lett.\ B {\bf 517},  233 (2001).


\bibitem{Ogul:2002ka}
  R.~Ogul and A.~S.~Botvina,
  Phys.\ Rev.\  C {\bf 66}, 051601 (2002).

\bibitem{Karnaukhov:03}
%
V.~A.~Karnaukhov {\it et al.},
  Phys.\ Rev.\  C {\bf 67},   011601 R (2003).
  
 \bibitem{Igor:2006xc} 
  A.~S.~Botvina  {\it et al.}, 
  Phys.\ Rev.\ C {\bf 74}, 044609 (2006)
 and references therein. 

\bibitem{Stanley:71}
%
see, for instance, 
H. E. Stanley, {\it Introduction to phase transitions and critical phenomena},
Clarendon Press, Oxford, 1971.

\bibitem{Bugaev:RVDW1}
%
K. A. Bugaev, M. I. Gorenstein, H. St\"ocker and W. Greiner,
Phys. Lett. B {\bf 485}, 121 (2000).


\bibitem{Bugaev:RVDW2}
%
G. Zeeb, K. A. Bugaev, P. T. Reuter and H. St\"ocker,
Ukr.  J. Phys.  {\bf 53},     279 (2008).

\bibitem{Bugaev:RVDW3}
%
K. A. Bugaev, 
 Nucl. Phys.  A  {\bf  807},    251  (2008).

  
\bibitem{Bugaev:07a}
K. A. Bugaev, 
Phys. Rev. {\bf C 76},    014903 (2007);
 Phys. Atom. Nucl.  {\bf 71}, 1615 (2008).
   
\bibitem{Bugaev:09a}
K. A. Bugaev, V. K. Petrov and G. M. Zinovjev,
Europhys. Lett. {\bf 85},  22002 (2009); 
Phys. Rev.  {\bf C  79},  054913    (2009).

\bibitem{Dillmann:91} 
%
A. Dillmann and G. E. Meier, 
J. Chem. Phys. {\bf 94}, 3872 (1991). 

\bibitem{LFK:94} 
%
A. Laaksonen, I. J. Ford, and M. Kulmala,
 Phys. Rev. E {\bf 49}, 5517 (1994).
 
 \bibitem{Hove}
L. Van Hove, Physica {\bf 15}, 951 (1949) ;   Physica {\bf 16},   137 (1950).

\bibitem{Fisher:70}
M. E. Fisher and B. U. Felderhof, Ann. of Phys. {\bf 58},  217 (1970).

\bibitem{Bugaev:Nucleation11}
%
K. A. Bugaev, A. I. Ivanitskii, E. G. Nikonov,  A. S. Sorin and G. M. Zinovjev,
Can We Rigorously Define Phases in a Finite System?, 
Chapter 18 of the Proceedings of the XV-th Research Workshop 
{\it ``Nucleation Theory and Applications"}, held at  JINR,  Dubna, Russia, April 1- 30, 2011, 
edited by J. W. P. Schmelzer, G. Ropke, V. B.  Priezzhev, Dubna JINR, 2011;
 arXiv:1106.5939 [nucl-th] 
 
\bibitem{Bugaev_09}
%
K. A. Bugaev, V. K. Petrov and G. M. Zinovjev, 
Phys.  Part. Nucl. Lett. {\bf 9},  238 (2012);
arXiv:0904.4420 [hep-ph] (2009).



\bibitem{Bugaev:04a} 
K. A. Bugaev, L. Phair and J. B. Elliott,
  Phys.\ Rev.\ E {\bf 72}, 047106 (2005).
  
   
\bibitem{Bugaev:07b} 
K. A. Bugaev  and J. B. Elliott,
Ukr. J. Phys. {\bf 52},  301 (2007).

\bibitem{Bugaev:12b}
%
A. I. Ivanytskyi, K. A. Bugaev,   A. S. Sorin and G. M. Zinovjev,
Phys. Rev. E {\bf 86},  061107 (2012).     
 


   
  
\bibitem{Satarov:2009zx} 
  L.~M.~Satarov, M.~N.~Dmitriev and I.~N.~Mishustin,
  Phys.\ Atom.\ Nucl.\  {\bf 72}, 1390 (2009).


    
\bibitem{Bugaev:2012c} 
  %
K. A. Bugaev et al.,
Phys. Atom. Nucl. {\bf 75},  707 (2012); 
arXiv:1101.4549 [hep-ph]
 
  


 
  
\bibitem{Bugaev:05}
%
K. A. Bugaev,
Acta. Phys. Polon. {\bf B 36}, 3083 (2005).
  
  

\bibitem{Kfactor:1}
%
D. Vretenar, T. Niksic, and P. Ring, Phys. Rev. C {\bf 68}, 024310 (2003). 


\bibitem{Kfactor:2}
%
G. Colo and Nguyen Van Giai, Nucl. Phys. A {\bf 731}, 15 (2004). 

\bibitem{Khan:2009}
%
E. Khan, Phys. Rev. C {\bf 80}, 011307(R) (2009). 

\bibitem{Kfactor:3}
%
V. B. Soubbotin, V. I. Tselyaev and  X. Vinas, 
Phys. Rev. C {\bf  69}, 064312 (2004). 

\bibitem{SMM:12}
%
N. Buyukcizmeci et al.,  arXiv:1211.5990v2  [nucl-th]. 


\bibitem{Campi:03}
%
X. Campi,  H. Krivine  E. Plagnol and N. Sator, 
Phys. Rev. C {\bf 67}, 044610 (2003).



\end{thebibliography}
\end{document}